\documentclass[a4paper,11pt]{article}
\pdfoutput=1

\usepackage[a4paper,
  left=2.5cm, right=2.5cm,
  top= 3cm, bottom=4cm]{geometry}

\usepackage{amsmath}
\numberwithin{equation}{section}

\usepackage{mathtools}
\usepackage{oldgerm}
\usepackage{amssymb}
\usepackage{bbm}
\usepackage{graphicx}
\usepackage{subcaption}
\usepackage{tikz-cd} 
\usetikzlibrary{backgrounds, arrows,calc,shapes,decorations.pathreplacing, decorations.markings, automata,positioning}
\usepackage{epic}
\usepackage{hyperref}
\usepackage{bbold}

\newcommand{\be}{\begin{equation}}  
\newcommand{\ee}{\end{equation}}  
\newcommand{\bp}{\begin{pmatrix*}[r]}  
\newcommand{\ep}{\end{pmatrix*}}  
\newcommand{\bpp}{\begin{pmatrix}}  
\newcommand{\epp}{\end{pmatrix}}  
\newcommand{\bcd}{\begin{center}
\begin{tikzcd}}
\newcommand{\ecd}{\end{tikzcd} \end{center}}

\tikzset{->-/.style={decoration={
  markings,
  mark=at position .5 with {\arrow{>}}},postaction={decorate}}}
  \pgfdeclareradialshading{ballshading}{\pgfpoint{-10bp}{10bp}}
 {color(0bp)=(gray!40!white); 
 color(9bp)=(gray!75!white);
 color(18bp)=(gray!70!black); 
 color(25bp)=(gray!50!black); 
 color(50bp)=(black)}

\def\R{\mathbb{R}}
\def\P{\mathbb{P}}
\def\cL{\mathcal{L}}
\def\C{\mathbb{C}}
\def\cO{\mathcal{O}}
\def\cS{\mathcal{S}}
\def\1{\mathbb{1}}
\def\cB{\mathcal{B}}

\begin{document}
\begin{titlepage}
 
\vskip -0.5cm
\rightline{\small{\tt  t14/168}} 
 
\begin{flushright}

\end{flushright}
 
\vskip 1cm
\begin{center}
 
{\huge \bf \boldmath F-theory on singular spaces} 
 
 \vskip 2cm
 
Andr\'es Collinucci$^1$ and Raffaele Savelli$^{2}$

 \vskip 0.4cm
 
 {\it  $^1$Physique Th\'eorique et Math\'ematique and International Solvay Institutes,\\ Universit\'e Libre de Bruxelles, C.P. 231, 1050
Bruxelles, Belgium \\[2mm]
 
 $^2$Institut de Physique Th\'eorique, CEA Saclay, Orme de Merisiers, F-91191 Gif-surYvette, France
 }
 \vskip 2cm
 
\abstract{We propose a framework for treating F-theory directly, without resolving or deforming its singularities. This allows us to explore new sectors of gauge theories, including exotic bound states such as T-branes, in a global context. We use the mathematical framework known as Eisenbud's matrix factorizations for hypersurface singularities.
We display the usefulness of this technique by way of examples, including affine singularities of both conifold and orbifold type, as well as a class of full-fledged compact elliptically fibered Calabi-Yau fourfolds.} 

\end{center}

\end{titlepage}

\tableofcontents

\section{Introduction}

\begin{figure}
\centering
\includegraphics[scale=.3]{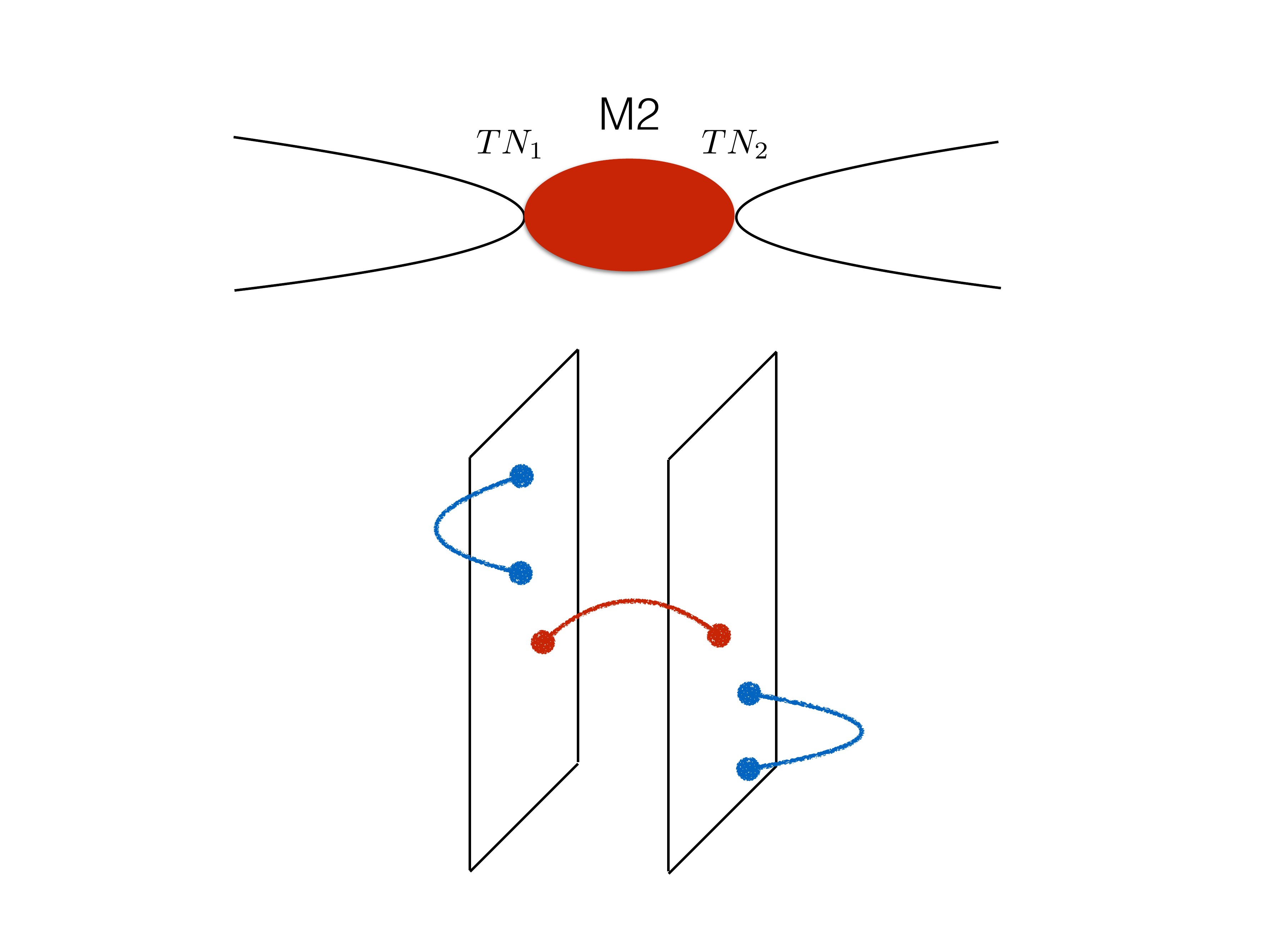}
\caption{Two parallel D6-branes. The blue strings uplift to moduli of two-centered Taub-NUT metric, whereas the red string uplifts to a `vanishing' M2-brane.}\label{D6M2Fig}
\end{figure}

F-theory \cite{Vafa:1996xn} is a geometric framework that describes both the gravitational as well as the gauge theory data of 7-branes in type IIB string theory. The target space of IIB string theory is combined with the data of the axio-dilaton into a unified twelve-dimensional space that is elliptically fibered over the standard ten-dimensional space. In situations of interest, this twelve-dimensional space is a (warped) product of $\mathbb{R}^{1,3}$ times an elliptically fibered Calabi-Yau (CY) fourfold. 
A priori, the elliptically fibered space only encodes the bulk supergravity data that backreacts to the presence of the 7-branes. However, when several 7-branes coincide, or intersect, extra massless degrees of freedom arise from open string (or string junction) excitations. In order to see this data in F-theory, it is most convenient to pass to the dual M-theory formulation. Via a chain of dualities, one ends up studying M-theory on $\mathbb{R}^{1,2}$ times the same CY fourfold. Now, those missing light degrees of freedom can be clearly accommodated as follows \cite{Witten:1995ex, Bershadsky:1995sp}:
When 7-branes coincide or intersect, the corresponding CY fourfold develops singularities. Such singular CY fourfold can usually be understood as limiting points in a family of smooth manifolds, whereby some two-dimensional submanifolds (spheres) are forced to shrink down to zero K\"ahler volume. Now, by postulating the existence of M2-branes in M-theory, light degrees of freedom arise as M2's wrapped on such vanishing spheres.
These degrees of freedom are crucial to make sense of the gauge theory data: They furnish fields both in the vector and chiral multiplets. In the case of coincident branes, they provide the root vectors of the non-Abelian Lie algebra. In the case of intersecting branes, they provide the bi-fundamental matter. In order to visualize this data, it is best to take one T-duality along the 7-brane worldvolume and work with D6-branes (see fig. \ref{D6M2Fig} for the case of coincident branes).

In perturbative string theory, open strings stretching from one brane to itself, or between two different branes can both be quantized with the same techniques, they are on equal footing. Yet, when lifted to M-theory, they become immensely different. The former typically lift to supergravity moduli of the 11d metric or the $C_3$-form, whereas the latter are accounted for by the presence of vanishing M2-branes. Because the latter require singularities in the geometry, any attempt to build a physically interesting model brings us to an {\it impasse}: To get any interesting physics in the effective field theory, we must force the CY fourfold to be singular. On the other hand, to get a sensible description of the microscopic theory, we must desingularize the CY fourfold.
The current \emph{modus operandi} in F-theory is roughly based on a two-step procedure \cite{Bershadsky:1996nh}:
\begin{enumerate}
\item Create an elliptically fibered fourfold over a particular K\"ahler threefold, and enforce a pattern of singularities by restricting its complex structure moduli. 
\item Desingularize the variety either via blow-ups, small resolutions, or deformations.
\end{enumerate}
The reasons for desingularizing are both technical as well as conceptual:
\begin{enumerate}
\item On a singular space, standard notions such as a metric or a 3-form are not well-defined. This makes a concrete description of the M-theory data difficult.
\item More fundamentally, the singularities of the 11d supergravity metric lead to singularities in the effective field theory, which are believed to be artifacts of incorrect Wilsonian reasoning. It is expected that M-theory has extra light degrees of freedom that have wrongly been integrated out. Incorporating them should cure such singularities, leading to a well-defined effective theory. The first example of such a phenomenon was seen in Strominger's treatment of the conifold in type IIB string theory \cite{Strominger:1995cz}, where a logarithmic divergence of the prepotential in $
\mathcal{N}=2, d=4$ is canceled by a one-loop contribution from light D3-particles wrapping a vanishing 3-sphere.
\end{enumerate}
By resolving or deforming the singularities, one not only recovers control over the geometry, but one also gains a better understanding of the singularities: One can keep track of which cycles will shrink to zero size upon blowing down or turning the deformation off. In this way, one hopes to still capture all essential data of the singular F-theory compactification. 
For instance, if one uplifts two intersecting D7-branes to F-theory, it is known that the CY fourfold develops a family of conifold singularities fibered over the type IIB matter curve. One can compute the chiral index of such matter by integrating the $G_4$ field-strength over the four-cycle that emanates from resolving this family. Furthermore, in \cite{Bies:2014sra}, the authors proposed and successfully implemented a formalism to compute the absolute spectrum (as opposed to an index) by treating the $C_3$-form as a Deligne cohomology class. So why should we worry about having to desingularize our space? The answer is two-fold:
\begin{enumerate}
\item The first objection to this procedure is philosophical: Resolving the singularities in the fourfold, seen as an M-theory compactification, corresponds to moving on a Coulomb branch in three dimensions. This, in turn, T-dualizes to turning on Wilson lines on the type IIB side that break 4d Poincar\'e invariance. Furthermore, upon taking the zero area limit for the elliptic fiber, the singularities resurrect, so in some sense, this branch is non-existent in F-theory.
The deformation picture avoids this problem, as it corresponds to moving on branches that do survive this F-theory limit \cite{Grassi:2013kha, Grassi:2014sda}.

\item The second objection has more drastic implications: When one resolves the singularities, one is rendering massive the light M2-branes associated with the vanishing cycles. In doing so, one can no longer switch on vev's for the effective fields describing them\footnote{Deforming has a more subtle effect. It corresponds essentially to either separating or joining the branes, thereby changing the basis of degrees of freedom. What would have corresponded to light stretched strings can become strings with both ends on the same brane.}
. An example of these fields are those corresponding to the bi-fundamental strings between two stacks of D7-branes. By resolving the family of conifold singularities which arises in M-theory, one is moving onto a Coulomb branch, thereby making \emph{inaccessible} the vacua where such fields acquire non-trivial expectation values.  Conversely, by switching on vev's for bifundamental matter, we are \emph{obstructing} the blow-up modes in M-theory.

\end{enumerate}
Therefore, by desingularizing the fourfold, one is ruling out a significant portion of possible supersymmetric backgrounds of the theory. In particular, it is not possible to access vacua that contain bound states of 7-branes, such as the ones dubbed `T-branes' \cite{Cecotti:2010bp}, and thus study their spectrum of massless fluctuations. This part of the spectrum was discussed in IIB string theory in \cite{Donagi:2011jy}, and suggestions were made as to its F-theory fate.

In \cite{Anderson:2013rka}, a new approach is proposed to deal with F-theory on singularities by first studying the deformed space, and carefully taking a singular limit that somehow retains the relevant structure to `see' the gluing degrees of freedom responsible for the bound states. The method involves the so-called \emph{mixed Hodge structures} of the deformed space, of which a singular limit is taken. What the authors observe is an emergent Hitchin structure that mimics the missing gauge theory data. At present, we do not know the relation between that strategy and the one we will present here. This is an interesting question to address in the future.

In this paper, we propose a new formalism that allows us to deal with singular varieties directly. It is based on the theory of \emph{matrix factorizations} invented by D. Eisenbud in 1980 \cite{Eisenbud:1980}. The underlying philosophy is the following: Instead of replacing our singular variety by a smooth one, we will define structures on the singularity that will give us a foothold on all the relevant holomorphic data. This is akin to what one does when probing a singularity in open string theory \cite{Douglas:1996sw, Berenstein:2001jr}: One supplements the singular geometry with `fractional branes' that enrich the coordinate ring of the space, and make it `smooth' in a very precise mathematical sense.

Matrix factorizations do not only allow us to calculate things directly on the singular space, but also allow us to construct the so-called \emph{non-commutative crepant resolutions} invented by Van Den Bergh \cite{Bergh:aa} (see \cite{Aspinwall:2010mw} for an accessible account), which, in many ways are better suited than traditional commutative resolutions. For example, in the case of a conifold-like singularity, one can desingularize the F-theory fourfold via small resolutions \cite{Esole:2011sm}. However, upon doing so, we are forced to choose among two small resolutions in a unnatural way. The two choices are related by a flop transition. From the 3d gauge theory point of view, this transition corresponds to a Weyl reflection on the enhanced $SU(2)$ which exchanges the two Coulomb branches of the theory \cite{Diaconescu:1998ua, Grimm:2011fx, Cvetic:2012xn, Hayashi:2013lra,Hayashi:2014kca}. The non-commutative crepant resolution is a description of the singularity that does not enforce such a choice. It exists precisely in the singular phase of the geometry, and, in a sense which can be made precise, contains both small resolutions in its entrails. 

Matrix factorizations have already made their appearance in string theory as a tool for encoding Landau-Ginzburg models with boundaries, see \cite{Kapustin:2002bi, Hori:2004zd, Brunner:2003dc} for some background, and \cite{Jockers:2007ng} for a review. Our application of this piece of mathematics will be completely different, as we will define matrix factorizations of the F-theory fourfold.

We will argue that the complete way to specify an F-theory compactification is to define a geometry, and a corresponding choice of matrix factorization. In \cite{Braun:2011zm} hints of this paradigm emerged, when the authors realized that, even for smooth F-theory compactifications, a matrix factorization of the fourfold corresponds to a choice of G-flux. Here, we will show that it encodes information about the non-Abelian degrees of freedom which are missing in the 11d supergravity approximation of M-theory. Moreover, it will allow us to explore the backgrounds which cannot be accessed by other techniques, and thus to study the spectra of massless fluctuations around them. We will describe a class of globally defined T-branes in section \ref{GlobalTbraneSec}, and explain in detail its peculiarities by using matrix factorizations.

This paper is organized as follows: In section \ref{MFreview} we will introduce the mathematical tools needed in the rest of the paper. In section \ref{ConifoldSection},
we will present our main results through three examples, all involving only abelian gauge symmetries. We will start with a six-dimensional toy model analyzed in affine space, and then turn to a four-dimensional one, where we introduce chirality while still working with a local CY fibration.
Our third example comprises a class of compact four-dimensional models obtained via the so-called `$U(1)$-restriction' \cite{Grimm:2010ez}. In section \ref{ResolutionSection}, we provide further motivation for the matrix factorizations picture of F-theory: We show how such a formal structure does not fall from the sky, but is intimately related to the geometry of the resolution of singularities, thus being a very appropriate candidate to analyze the physics hidden in them. In section \ref{BROBsection}, we will describe how this formalism is able to deal with the Higgsing of non-abelian gauge theories, focusing on the case of $SU(n)$ singularities. Section \ref{ConclSection} contains our concluding remarks and speculations.

\section{Introduction to matrix factorizations}\label{MFreview}

\subsection{Basic definitions}\label{DefSection}

Let us introduce the concept of matrix factorizations \cite{Eisenbud:1980}, which is central to our proposal. See \cite{Yoshino, MCMbook} for some mathematical background, and \cite{Hori:2004zd, Herbst:2008jq} for other physical applications. The definition is astonishingly simple, but has deep connections to modern algebraic geometry. Matrix factorizations are a mathematical device that probes the detailed structure of hypersurface singularities. 
In what follows, we will try to strike a balance between legibility and mathematical precision, by introducing notions on a need-to-know basis.
The basic idea is very simple. Given a polynomial $P$ of some coordinate ring in an affine space, a matrix factorization is a pair $(A, B)$ of square matrices such that
\be 
A \cdot B = B \cdot A = \mathbb{1} \cdot P\,.
\ee
A given polynomial can admit a host of matrix factorizations of arbitrary size, but in certain situations one can classify \emph{irreducible} matrix factorizations (MF's for convenience), which serve as building blocks for other MF's. For instance, given two MF's $(A_1, B_1), (A_2, B_2)$, we can define the direct sum as
\be
(A_1, B_1) \oplus (A_2, B_2) \equiv \left( \bpp A_1 & 0 \\ 0 & A_2 \epp, \bpp B_1 & 0 \\ 0 & B_2 \epp \right)\,.
\ee
Clearly, the following two MF's are not very interesting: $(1, P)$ and $(P, 1)$. They always exist, and give no extra information about the singularity. In what follows, we refer to them as the  \emph{trivial MF} and the \emph{non-reduced MF}, respectively. In fact, any MF that contains $(P, 1)$ as a direct summand is referred to as \emph{non-reduced} or \emph{non-stable}.
In affine space, a hypersurface equation admits a number of \emph{reduced, non-trivial} MF's if and only if it is singular. Each of them gives us some information about the structure of the singularity. 

The simplest case is when a singularity admits a $1 \times 1$ MF. Suppose that $P$ is given by
\be 
P \equiv P_1 \cdot P_2 
\ee
for two generic polynomials $P_1$ and $P_2$. In this case there are at least two inequivalent, non-trivial MF's, $(P_1, P_2)$ and $(P_2, P_1)$. What these MF's are telling us, is that our variety has two components. 
A more interesting yet familiar situation arises in the case of the conifold
\be\label{ConifoldPolynomial}
P\equiv x\,y+u\,v  \quad \in \C[u,v,x,y]\,,
\ee
which admits two irreducible MF's up to base redefinitions:
\be
(\phi, \psi) \quad \text{and} \quad (\psi, \phi)\,.
\ee
for
\be
\phi \equiv \bp x & -u \\ v & y \ep, \quad {\rm and} \quad \psi \equiv \bp y & u \\ -v & x \ep \,.
\ee
The first MF is telling us that the conifold has a family of non-Cartier divisors. These are given by ideals defined through Im$(\phi)$, i.e. by the loci
\be\label{non-Cartier-div}
a\,x-b\,u = 0 \quad \cap \quad a\,v+b\,y=0 \quad \subset \C^4\,,
\ee
where $a,b$ are complex numbers. In fact, since $a$ and $b$ are defined modulo rescaling, and we exclude $a=b=0$, they can be thought of as the homogeneous coordinates of a $\P^1$. Analogously, the second MF tells us that there is a second such family of non-Cartier divisors at ideals defined by Im$(\psi)$. One can verify that the intersection \eqref{non-Cartier-div} is automatically contained in the threefold defined by the zero-locus of \eqref{ConifoldPolynomial}. A non-Cartier divisor is an instance of a codimension one \emph{non-regular subscheme}. Roughly, a codimension $d$ subscheme is said to be regular if it is locally a complete intersection of $d$ hypersurfaces with the variety in question.
A singularity that admits irreducible MF's of size bigger than $2$ by $2$ has higher codimension non-regular subschemes.

\subsection{Relation to D7-branes}\label{RelationToD7s}

What is the relation of all this to D7-branes? In a companion paper \cite{Collinucci:2014qfa} we discussed D-branes from the perspective of \emph{tachyon condensation} \cite{Sen:1998sm}. In this context, D-branes are viewed as \emph{complexes of vector bundles}. A complex of vector bundles is a collection\footnote{For the purposes of this paper it suffices to focus on finite collections.} of vector bundles $\{A_i\}_i$ and maps between them $\{d_i\}_i$
\be
\begin{tikzcd}
A_\bullet: & A_1 \rar{d_1} & A_{2} \rar{d_{2}} & \ldots \rar{d_{n-1}} & A_n\,,
\end{tikzcd}
\ee
such that $d_{i-1} \circ d_i = 0$. 
A map $m_\bullet$ between two complexes $A_\bullet$ and $B_\bullet$, called cochain map, is a collection of maps $\{m_i\}_i$ such that all squares of the following diagram commute
\be\label{cochainmapdef}
\begin{tikzcd}[row sep=huge, column sep=large]
A_1  \dar{m_1} \rar{d_1^A} & A_{2} \dar{m_2} \rar{d_2^A} & \ldots \rar{d_{n-2}^A} & A_{n-1} \dar{m_{n-1}} \rar{d_{n-1}^A} & A_n \dar{m_n}  \\
B_1 \rar{d_1^B} & B_2 \rar{d_2^B}  & \ldots \rar{d_{n-2}^B} & B_{n-1}  \rar{d_{n-1}^B} & B_n 
\end{tikzcd}
\ee
These maps are defined modulo the so-called \emph{homotopies}: A cochain map $m_{\bullet}$ is declared to be zero if there are diagonal maps $\{h_i\}_i$ in
\be\label{homotopydef}
\begin{tikzcd}[row sep=huge, column sep=large]
A_1  \dar{m_1} \rar{d_1^A} & A_{2} \dar{m_2} \rar{d_2^A} \dlar[near start, dashed]{h_1} & \ldots \rar{d_{n-2}^A} \dlar[near start, dashed]{h_2} & A_{n-1} \dar{m_{n-1}} \rar{d_{n-1}^A} \dlar[near start, dashed]{h_{n-2}} & A_n \dar{m_n} \dlar[near start, dashed]{h_{n-1}} \\
B_1 \rar{d_1^B} & B_2 \rar{d_2^B}  & \ldots \rar{d_{n-2}^B} & B_{n-1}  \rar{d_{n-1}^B} & B_n 
\end{tikzcd}
\ee
such that $m_i = d_{i-1}^B \circ h_{i-1}+ h_{i} \circ d^A_{i}\;\forall i$.

A D7-brane is defined by a two-term complex of the form:
\be
\begin{tikzcd}
E \rar{T}& F\,,
\end{tikzcd}
\ee
such that the cokernel sheaf $\cS=$coker$(T)$ of the `tachyon' map $T$ has support only over the hypersurface $P_{D7}=0$ wrapped by the D7-brane. If we would multiply a section $s$ of the sheaf $\cS$ by $P_{D7}$, then necessarily $s \cdot P_{D7} = 0$. In other words, a wave-function that is localized on the D7-brane is necessarily annihilated by the polynomial that vanishes on said brane. Let us carry out this multiplication at the level of the complex:
\be
\begin{tikzcd}[column sep=huge]
E \dar{P_{D7}} \rar{T}& F \dar{P_{D7}}\\
E \rar{T}& F 
\end{tikzcd}
\ee
Since $P_{D7}$ annihilates the cokernel $\cS$ of the complex, then this cochain map should be equivalent to multiplying by zero \emph{up to homotopy}. Therefore, there must exist a contracting homotopy
\be
\begin{tikzcd}[column sep=huge, row sep=huge]
E \dar{P_{D7}} \rar{T}& F \dar{P_{D7}} \dlar[dashed, near start]{\tilde T}\\
E \rar{T}& F 
\end{tikzcd}
\ee
such that both vertical maps can be gauged away, i.e. such that $T \cdot \tilde T = \tilde T \cdot T = P_{D7} \cdot \mathbb{1}$. 
\emph{Therefore, whenever we discuss D7-branes as tachyon condensates between two stacks of D9's and anti-D9's, we are building a matrix factorization of the hypersurface equation of the D7-brane.}

We would like to stress that this is \emph{not} the treatment of D-branes via matrix factorizations that has appeared in the Landau-Ginzburg literature \cite{Kapustin:2002bi, Hori:2004zd, Brunner:2003dc}. In that case, one considers MF's for the hypersurface equation defining a CY threefold in which D-branes live. Here, we are constructing MF's for the equation defining the D7-brane itself.

\subsection{Kn\"orrer's periodicity}\label{KnoerrerSection}
In \cite{Collinucci:2014qfa} we have developed tools to treat D7-branes in type IIB string theory and we have just seen that we have implicitly constructed MF's for the hypersurfaces wrapped by them. In this section, we will introduce a correspondence of categories of MF's that will serve as our prototype duality between type IIB and F-theory. We will use this correspondence just to lay down some abstract technology, which we will then extrapolate and adopt to analyze general F-theory backgrounds without any reference to type IIB.

In 1987, Horst Kn\"orrer \cite{Knorrer} proved that, under special conditions, two hypersurfaces describing completely different spaces could have equivalent sets of matrix factorizations.
Given a polynomial $P$ in some ring $S$, augment $S$ by two coordinates $S[u, v]$. This amounts to taking the ring $S$ and throwing in two extra coordinates. Now define the new hypersurface $P+u v$. Kn\"orrer proved that
\be
\text{MF's of} \qquad P \in S  \qquad \longleftrightarrow \qquad \text{MF's of} \quad P+u v \in S[u,v]\,.
\ee
In order to clarify what we mean by the deliberately vague symbol ``$\longleftrightarrow$", we must define some structures on the set of matrix factorizations that promote it to a category. Fix a ring $S$ for an affine space, and a polynomial $P \in S$. Let MF$(P)$ be the set of all MF's of $P$. Let $(\phi_1, \psi_1)$ and $(\phi_2, \psi_2)$ be two elements of MF$(P)$ of sizes $n_1 \times n_1$ and $n_2 \times n_2$, respectively. Then, we define a \emph{morphism} between them as a pair of maps $(\alpha, \beta): (\phi_1, \psi_1) \rightarrow (\phi_2, \psi_2)$ such that the following squares commute:
\be\label{MFmorphisms}
\begin{tikzcd}
S^{\oplus n_1} \rar{\psi_1} \dar{\alpha} & S^{\oplus n_1} \dar{\beta} \\
S^{\oplus n_2} \rar{\psi_2} & S^{\oplus n_2} 
\end{tikzcd}
\qquad\quad 
\begin{tikzcd}
S^{\oplus n_1}\rar{\phi_1} \dar{\beta} & S^{\oplus n_1} \dar{\alpha}\\ S^{\oplus n_2} \rar{\phi_2} & S^{\oplus n_2}
\end{tikzcd}
\ee
i.e. such that 
\be
\alpha \circ \phi_1 = \phi_2 \circ \beta \qquad {\rm and} \qquad \beta \circ \psi_1 = \psi_2 \circ \alpha \,.
\ee
In fact, it is easy to realize that either one of the above conditions implies the other, and thus it suffices to consider only one of the squares in \eqref{MFmorphisms}.
These morphisms give MF$(P)$ the structure of a category. 

Just as we can define objects as kernels or cokernels of maps between sheaves, so can we use morphisms between MF's to define other objects, via the so-called \emph{cone construction}. 
Given two complexes $A_\bullet$, $B_\bullet$, and a cochain map $m_\bullet$ between them, as in \eqref{cochainmapdef}, we define the \emph{mapping cone} of $m_\bullet$ as the following complex
\begin{equation}
\begin{tikzpicture}[baseline=(current bounding box.center), row sep=3em, column sep=2.8em, text height=1.5ex, text depth=0.25ex]
\node (upleft) at (-1.5,0) {$\cdots$};
\node (downleft) at (-1.5, -1.4) {$\cdots$};
\node (upright) at (7.5,0) {$\cdots$};
\node (downright) at (7.5,-1.4) {$\cdots$};
\node (A) at (0,0) {$A_{i}$};
\node (B) at (3,0) {$A_{i+1}$};
\node (C) at (6,0) {$A_{i+2}$};
\node (p1) at (0,-.7) {$\oplus$};
\node (p2) at (3,-.7) {$\oplus$};
\node (p3) at (6,-.7) {$\oplus$};
\node (D) at (0,-1.4) {$B_{i-1}$};
\node (E) at (3,-1.4) {$B_i$};
\node (F) at (6,-1.4) {$B_{i+1}$};
\path[->,font=\scriptsize]
(A) edge node[auto] {$-d^A_{i}$} (B)
(B) edge node[auto] {$-d^A_{i+1}$} (C)
(D) edge node[auto] {$d^B_{i-1}$} (E)
(E) edge node[auto] {$d^B_{i}$} (F)
(A) edge node[above] {$m_i$} (E)
(B) edge node[above] {$m_{i+1}$} (F)
(upleft) edge (A)
(downleft) edge (D)
(C) edge (upright)
(F) edge (downright)
;
\end{tikzpicture}
\end{equation}
Now, as explained in section \ref{DefSection}, the MF given by $(1, P)$ is considered trivial, as it corresponds to the sheaf
\be
\begin{tikzcd}
S \rar{1} & S\,,
\end{tikzcd}
\ee
which has trivial cokernel. On the other hand, the MF $(P, 1)$, although not trivial, is still uninteresting, since it does not in any way `probe' the singular structure of our space. It just gives us the coordinate ring of the hypersurface via the exact sequence
\be
\begin{tikzcd}
0 \rar & S \rar{P} & S \rar & S/(P) \rar & 0\,.
\end{tikzcd}
\ee
Throughout the paper, we will often work in the \emph{stable category} \underline{MF}$(P)$, which is defined as MF$(P)$, modulo all morphisms that factor through finite direct sums of $(P,1)$. For example, take a size $n$ MF $(\phi, \psi) \in {\rm MF}(P)$, and $n$ copies of $(P, 1)$ and define the following morphism\footnote{This is a cochain map between two complexes and corresponds to just the right-hand square of \eqref{MFmorphisms}.}:
\be
\begin{tikzcd}[column sep=huge]
S^{\oplus n} \rar{P \cdot \mathbb{1}_n} \dar{\psi} & S^{\oplus n} \dar{\mathbb{1}_n} \\
S^{\oplus n} \rar{\phi} & S^{\oplus n}
\end{tikzcd}
\ee
If we now take the mapping cone of such a morphism, we obtain the following complex
\be\label{ConeBeforeChange}
\begin{tikzcd}[column sep=huge]
S^{\oplus n} \rar{\bp -P \\ \psi \ep} & S^{\oplus 2n} \rar{(\mathbb{1}_n\,,\, \phi)} & \underline{S^{\oplus n}}\,.
\end{tikzcd}
\ee
Here, we have underlined an object to denote it as the starting zeroth position in the complex.
This complex is equivalent to $(\psi, \phi)$ shifted one place to the left, i.e.
\be\label{ConeAfterChange}
\begin{tikzcd}
S^{\oplus n} \rar{\psi} & S^{\oplus n} \rar & \underline{0}\,,
\end{tikzcd}
\ee
as can be easily seen by first performing on \eqref{ConeBeforeChange} the automorphism defined by the following cochain map
\begin{equation}
\begin{tikzcd}[column sep=80pt,row sep=35pt]
S^{\oplus n} \arrow{r}{\bp -P \\ \psi \ep} \arrow{d}{\1_n} & S^{\oplus 2n}  \arrow{r}{(\mathbb{1}_n\,,\, \phi)} \arrow{d}{\mathbb{a}} & \underline{S^{\oplus n} \arrow{d}{\1_n}}\\
S^{\oplus n} \arrow{r}{\bp 0 \\ \psi \ep} & S^{\oplus 2n}  \arrow{r}{(\mathbb{1}_n\, ,\, 0)} & \underline{S^{\oplus n}}
\end{tikzcd}\quad\qquad \mathbb{a}=\begin{pmatrix} \mathbb{1}_n & \phi\\ 0 & \mathbb{1}_n \end{pmatrix}\,,
\end{equation}
and then discarding the trivial complex.
We will refer to the complex \eqref{ConeAfterChange} as $(\psi, \phi)[1]$. Since this new complex was obtained as the cone of a morphism from $(P, 1)$ to $(\phi, \psi)$, we can say that, in the stable category \underline{MF}$(P)$, we have
\be\label{StableCatEquiv}
(\phi, \psi) \cong (\psi, \phi)[1]\,.
\ee

We are now ready to formulate the theorem known as Kn\"orrer's periodicity: 
\be
\text{\underline{MF}}(P) \quad  \text{for} \quad P \in S \qquad \cong \qquad \text{\underline{MF}}(P+u v) \quad \text{for} \quad P+u v \in S[u,v]\,.
\ee
The easy part of this statement is the explicit construction of an MF for $P+ uv$, given an MF $(\phi, \psi)$ of size $n$ for $P$. It is simply an MF of size $2 n$ given by
\be \label{upstairs}
\left( \bpp \phi & -u \cdot \1_n \\ v \cdot \1_n & \psi \epp\,,\, \bpp \psi & u \cdot \1_n \\ -v \cdot \1_n & \phi \epp \right)\,.
\ee
The other way around is less straightforward. Given an MF $(\Phi, \Psi)$ for $P+ u v$, it turns out that it becomes reducible if one sets the variables $u$ and $v$ to zero in all entries, and the reducible components are such that
\be\label{u=v=0}
\Phi |_{u=v=0} \cong \bpp \phi & 0 \\ 0 & \psi \epp\,, \qquad \Psi |_{u=v=0} \cong \bpp \psi & 0 \\ 0 & \phi \epp\,,
\ee
for some $(\phi, \psi) \in \underline{\rm MF}(P)$. The most non-trivial part of the correspondence concerns the morphisms. On face value, two MF's in MF$(P+u v)$ will admit more morphisms between them than their dimensionally reduced counterparts in MF$(P)$. However, when we work in the respective \emph{stable} categories, we are modding out by a lot of morphisms. It is this crucial fact that makes Kn\"orrer's periodicity possible. Explicitly, given a morphism between two MF's of $P$, $(\alpha, \beta): (\phi_1, \psi_1) \rightarrow (\phi_2, \psi_2)$, its lift to a morphism between the corresponding MF's of $P+uv$ is 
\be\label{KnoerrerMorphisms}
\left( \bpp \alpha & 0 \\ 0& \beta \epp\,, \bpp \beta & 0 \\ 0  & \alpha \epp \right)\,:\qquad (\Phi_1\,,\,\Psi_1)\longrightarrow(\Phi_2\,,\,\Psi_2)\;,
\ee 
where $(\Phi_1,\Psi_1)$ and $(\Phi_2,\Psi_2)$ are constructed respectively from $(\phi_1, \psi_1)$ and $(\phi_2, \psi_2)$ as in \eqref{upstairs}. The astonishing result of Kn\"orrer is that every morphism $(\Phi_1\,,\,\Psi_1)\rightarrow(\Phi_2\,,\,\Psi_2) \in$ \underline{MF}$(P+u v)$ is of the form \eqref{KnoerrerMorphisms}.

\section{F-theory on singularities} \label{ConifoldSection}

Having introduced the framework of matrix factorizations for hypersurface singularities, we are now ready to make our proposal for computing spectra in F-theory. The strategy is as follows: 
Given a singular F-theory geometry defined by the zero locus of a (Weierstrass) polynomial $P$, classify all possible irreducible matrix factorizations of $P$. In order to fully specify an F-theory background, one chooses a specific $(\phi, \psi) \in$ MF$(P)$ as the extra required data. Part of the spectrum of this background is given by the supergravity moduli of the CY fourfold. The other part of the spectrum is given by the light M2-branes wrapping vanishing cycles. In order to find the latter, we compute the $\mathcal{E}xt^1({\rm coker}\phi, {\rm coker}\phi)$ sheaf in the stable category \underline{MF}$(P)$. The sections of these sheaves will correspond to the matter fields of interest. In cases where matter is expected to be localized on a curve $\mathcal{C}$, we will indeed find that this sheaf has support on $\mathcal{C}$. In cases with point-like matter, we will find that the sheaf is a skyscraper over points. 

This strategy is designed so that one is working directly on the singular F-theory fourfold. In order to test our proposal, we will study a case where we have complete control in perturbative type IIB string theory, and then redefine it directly in F-theory. In sections \ref{ConifoldAffine} and \ref{4dchirality} we will start on the type IIB side, and exploit {\it Kn\"orrer's periodicity} to map the data to F-theory. In sections \ref{U1restrictionSection} and \ref{GlobalTbraneSec}, on the contrary, we apply directly our technique to a class of compact F-theory fourfolds, and only afterwards compare the result to the type IIB expectations.

\subsection{The affine case}\label{ConifoldAffine}

\begin{figure}
\centering
\includegraphics[scale=.3]{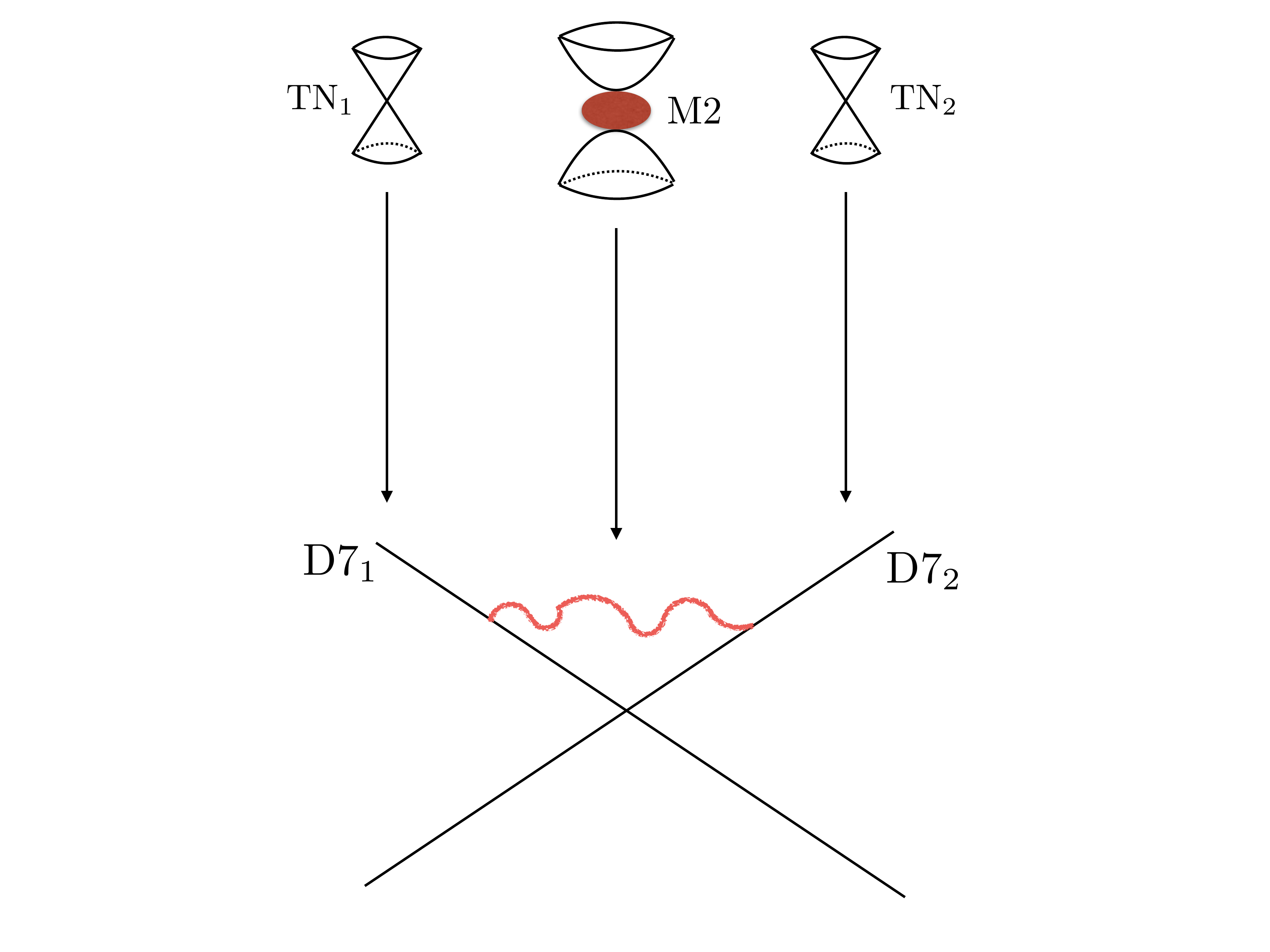}
\caption{Two intersecting D7-branes lifting to a conifold geometry, resulting from the collision of two families of Taub-NUT spaces. The red string uplifts to a `vanishing' M2-brane at the tip of the conifold.}\label{D7intM2Fig}
\end{figure}

Take type IIB string theory on $\C^2 \times \R^{1,5}$, with $\C[z_1, z_2]$ the coordinate ring of the `internal' $\C^2$, and place two intersecting D7-branes at $z_1=0$ and $z_2=0$. We are going to discard the factor $\R^{1,5}$, which is just a spectator. From the tachyon condensation perspective, the combined system is given by the cokernel sheaf of the following map \cite{Collinucci:2014qfa}
\be\label{2BraneSystem6d}
\begin{tikzcd}[column sep=large, ampersand replacement=\&]
S^{\oplus 2} \rar{\bpp z_1 & 0 \\ 0 & z_2 \epp} \& S^{\oplus 2}
\end{tikzcd}
\ee
which is a sheaf with support over the ideal $(z_1 \cdot z_2)$. This complex corresponds to a reducible matrix factorization of the polynomial $z_1 \cdot z_2$, which we can write as
\be \label{mfdownstairs}
(z_1, z_2) \oplus (z_2, z_1) \qquad {\rm in} \quad {\rm MF}(z_1 \cdot z_2)\,.
\ee
What is the F-theory lift of this configuration? It is given by an elliptic fibration over $\C^2$ defined as the following hypersurface:
\be
Y^2 = X^3+X^2\, Z^2-z_1 \, z_2\, \, Z^6 \qquad \subset \quad \C^2 \times \P^2_{2, 3, 1}\,,
\ee
where the first $\C^2$ is the base of the fibration, and the weighted projective space has homogeneous coordinates $[X: Y: Z]$. This space is singular at the codimension three locus given by the ideal $(Y, X, z_1, z_2)$. Since $Z$ cannot vanish at that locus, we fix its value to $Z=1$ with the available projective rescaling. Also, we can focus on the vicinity of the singularity and drop the $X^3$ term. This decompactifies the fiber and is equivalent to taking a weak coupling limit as defined in \cite{Clingher:2012rg,Donagi:2012ts} (see also \cite{Braun:2014nva}). Defining new coordinates $u$ and $v$ as $Y\pm X$, we find
\be
u\, v + z_1\, z_2 \subset \C[z_1, z_2, u, v]\,,
\ee
which defines a CY threefold with conifold geometry. The situation is summarized in fig. \ref{D7intM2Fig}.

 Now we can exploit Kn\"orrer's periodicity, and uplift the MF \eqref{mfdownstairs} to an MF of our F-theory threefold. Using formula \eqref{upstairs}, we find after suitable base transformations
\be\label{mfupstairsRed}
(\phi, \psi) \oplus (\psi, \phi)\,,
\ee
with
\be \label{phipsi}
\phi \equiv \bp z_1 & -u \\ v & z_2 \ep, \quad {\rm and} \quad \psi \equiv \bp z_2 & u \\ -v & z_1 \ep \,.
\ee
What does this reducible MF represent physically? Its two components are respectively the lift of the two components of \eqref{mfdownstairs}, which in turn are associated to the two intersecting D7-branes. The MF \eqref{mfupstairsRed} is pointing out the two families of non-Cartier divisors of the CY threefold, which in turn are in correspondence with two cohomology classes of two-forms that are only present due to the singularity. These two classes give rise to the $U(1)$ gauge fields living on the two D7-branes, by reducing the supergravity $C_3$ form along them. This is in complete agreement with the IIB expectation that there be a $U(1) \times U(1)$ gauge group, modulo mechanisms that might render photons massive such as the ones explored in \cite{Grimm:2010ez}.

Suppose now that we turn on an off-diagonal vev in our tachyon
\be
T = \bpp z_1 & 1 \\ 0 & z_2 \epp\,.
\ee
We know from section 3 of \cite{Collinucci:2014qfa} that we can make a base transformation that sends it to
\be 
T \longrightarrow T' = \bpp z_1 z_2 & 0 \\ 0 & 1 \epp \cong z_1\,z_2\,,
\ee
where, in the last step, we have eliminated a trivial brane/anti-brane pair. By Kn\"orrer's periodicity, we can use formula \eqref{upstairs} to uplift this new tachyon to a new MF of the F-theory threefold. After suitable exchanges of rows and columns, we find
\be \label{mfupstairs}
\left( \bpp \phi & \1_2 \\ 0 & \psi \epp, \bpp \psi & -\1_2 \\ 0 & \phi \epp \right)\,,
\ee
with $\phi$ and $\psi$ as in \eqref{phipsi}. Focusing on the first matrix of the MF, it is a simple matter to perform a  base transformation
\be\label{BasisChangeC}
\begin{tikzpicture}[scale=1.5]
\node (A) at (0,0) {$S^{\oplus 4}$};
\node (B) at (2,0)  {$S^{\oplus 4}$};
 \draw (A) edge[->] node[above, font=\scriptsize]{$\bpp \phi & \1_2 \\ 0 & \psi \epp$} (B);
 \draw(A) edge[loop below] node[font=\scriptsize]{$\begin{pmatrix} \mathbb{1}_2 & 0\\ -\phi & \mathbb{1}_2 \end{pmatrix}$} (A) ;
 \draw(B) edge[loop below] node[font=\scriptsize]{$\begin{pmatrix} \psi & -\mathbb{1}_2 \\ \1_2 & 0 \end{pmatrix}$} (B) ;
\end{tikzpicture} 
\ee
which, after discarding a trivial summand, yields
\be
\begin{tikzcd}[ampersand replacement=\&, column sep=large]
S^{\oplus 2} \rar{P \cdot \1_2} \& S^{\oplus 2}\,.
\end{tikzcd}
\ee
One can similarly track how the second matrix in \eqref{mfupstairs} behaves under this transformation. In the end, we find
\be
\left( \bpp \phi & \1_2 \\ 0 & \psi \epp, \bpp \psi & -\1_2 \\ 0 & \phi \epp \right) \quad \cong \quad (P \cdot \1_2\,,\, \1_2 ) \quad \cong \quad 0 \quad \in {\rm \underline{MF}}(uv+z_1 z_2)\,.
\ee
The conclusion is that, by switching on the off-diagonal term in the MF, we have `eaten up' all the information about the non-Cartier divisors. Therefore, our new MF  \eqref{mfupstairs} no longer tells us about the presence of two independent $U(1)$'s. It just  keeps track of the center of mass $U(1)$, which thus remains unbroken but has nothing to do with the still singular structure of the threefold geometry.

This is the picture we propose for doing F-theory: An F-theory background is not only specified by hypersurface polynomial plus $C_3$-form. One must supplement this information by \emph{a choice of matrix factorization} of the CY fibration. This MF will tell us which gauge groups are really present, and which ones are broken.

\subsection{Chiral gluing modes}\label{4dchirality}

The previous section contained an example of two D7-branes intersecting over six non-compact dimensions. We would now like to see a case where two D7-branes intersect over a compact Riemann surface, giving rise to a chiral spectrum in four dimensions. The simplest setup for this is to take type IIB string theory on $\tilde X \times \R^{1,3}$, where $\tilde X$ is the resolved conifold, that contains a $\P^1$ over which the branes can intersect. Again we will be discarding the irrelevant factor $\R^{1,3}$. Let us define our type IIB conifold as the toric space
\be
\tilde{X}:\quad\begin{tabular}{c|c|c|c}
$\sigma_1$ & $\sigma_2$ & $z_1$ & $z_2$ \\
\hline
$1$ & $1$ & $-1$ & $-1$
\end{tabular}
\ee
and choose the K\"ahler cone such that $\sigma_1$ and $\sigma_2$ parametrize a $\P^1$. Now let us place a D7-brane on $z_1=0$, with flux given by the line bundle $\cO(n_1)$, and one at $z_2=0$ with line bundle $\cO(n_2)$. Their intersection is the $\P^1$, hence we expect there to be a finite-dimensional chiral spectrum of trapped bifundamental strings.

Like in \eqref{2BraneSystem6d}, the system with two branes, denoted $\cB_i$ for $i=1, 2$, is represented by the direct sum of two complexes as follows
\be
\begin{tikzcd}[ampersand replacement=\&, column sep =huge] \label{tachyonresoconifold}
\begin{matrix} \cO(n_1+1) \\ \oplus \\ \cO(n_2+1)  \end{matrix} \rar{\bpp z_1 & 0 \\ 0 & z_2 \epp} \& \begin{matrix} \cO(n_1) \\ \oplus \\ \cO(n_2)  \end{matrix}
\end{tikzcd}
\ee
\vskip .2cm
\noindent The tachyon map is the first matrix of the factorization
\be \label{mfresconifold}
(z_1, z_2)\oplus (z_2, z_1) \quad \cong \quad \left( \bpp z_1 & 0 \\ 0 & z_2 \epp, \bpp z_2 & 0 \\ 0 & z_1 \epp \right) \quad \in \quad {\rm MF}(z_1\cdot z_2)\,.
\ee
The chiral spectrum, given by Ext$^1(\cB_2, \cB_1)$, as found in \cite{Katz:2002gh}, is easily computed in the derived category as Hom$(\cB_2, \cB_1[1])$ (see \cite{Collinucci:2014qfa} for a summary of these concepts, and \cite{Herbst:2008jq} for more detailed explanations). Concretely, it is given by the set of vertical maps $\varphi$ in the following diagram:
\be\label{Ext1-4d-D7}
\begin{tikzcd}[column sep=large, row sep=large]
& \cO(n_2+1) \dlar[dashed, near start]{h_L} \rar{\cdot z_2} \dar{\varphi} & \cO(n_2) \dlar[dashed, near start]{h_R} \\
\cO(n_1+1) \rar{\cdot z_1} & \cO(n_1)
\end{tikzcd}
\ee
modulo homotopies, i.e. $\varphi \sim \varphi +z_1\,h_L+h_R\,z_2$. The homotopies mod out the ideal $(z_1, z_2)$, thereby localizing the modes of $\varphi$ to the $\P^1$, as expected. Hence, we conclude that
\be \label{iibext}
{\rm Ext}^1(\cB_2, \cB_1)  \quad \cong \quad H^0(\P^1, \cO(n_1-n_2-1)) \quad \cong \quad \begin{cases} \C^{n_1-n_2} \quad {\rm for} \quad n_1> n_2 \\
0 \quad {\rm for} \quad n_1 \leq n_2 \,.\end{cases}
\ee
The shift by $-1$ in the resulting line bundle, which arises from the degrees of \eqref{Ext1-4d-D7}, correctly accounts for the $-c_1(\P^1)/2$ shift due to the Freed-Witten anomaly \cite{Freed:1999vc}.
With more effort, or by using a spectral sequence, one can show that 
\be
{\rm Ext}^2(\cB_2, \cB_1) \quad \cong \quad \begin{cases} \C^{n_2-n_1} \quad {\rm for} \quad n_2> n_1 \\
0 \quad {\rm for} \quad n_2 \leq n_1\,. \end{cases}
\ee
A faster way to arrive at this result goes as follows: The tachyon in \eqref{tachyonresoconifold} represents a direct sum of two sheaves, which can be represented as the trivial extension sequence:
\be
\begin{tikzcd}
0 \rar & \cB_1 \rar & \cB_1 \oplus \cB_2 \rar & \cB_2 \rar & 0\,.
\end{tikzcd}
\ee
This sequence is the trivial element in Ext$^1(\cB_2, \cB_1)$. A non-trivial element would correspond to a bound state of $\cB_1$ and $\cB_2$ that is \emph{not a direct sum}. Switching on an off-diagonal element in \eqref{tachyonresoconifold} accomplishes precisely that. The following sequence of complexes illustrates this:
\be
\begin{tikzpicture}[baseline=(current bounding box.center), row sep=3em, column sep=.5em, text height=1.5ex, text depth=0.25ex]
\node (A) at (0,0) {$\cO(n_1+1)$};
\node (B) at (4,0)  {$\cO(n_1)$};
\node (C) at (0,-2.5)  {$\cO(n_1+1)$};
\node (D) at (4,-2.5)  {$\cO(n_1)$};
\node[minimum size=4.7em] (E) at (0,-3)  {$\oplus$};
\node[minimum size=4.7em] (F) at (4,-3)  {$\oplus$};
\node (G) at (0,-3.5)  {$\cO(n_2+1)$};
\node (H) at (4,-3.5)  {$\cO(n_2)$};
\node (I) at (0,-6)  {$\cO(n_2+1)$};
\node (J) at (4,-6)  {$\cO(n_2)$};
\node (AA) at (8,0) {$\cB_1$};
\node (BB) at (8,-3) {bound state};
\node (CC) at (8,-6) {$\cB_2$};

\path[->,font=\scriptsize]
(A) edge node[auto] {$\cdot z_1$} (B)
(A) edge[red] node[black, auto] {$\bpp 1 \\ 0 \epp$} (C)
(G) edge[blue] node[black, auto] {$\bp 0 & 1 \ep$} (I)
(B) edge[red] node[black, auto] {$\bpp 1 \\ 0 \epp$} (D)
(H) edge[blue] node[black, auto] {$\bp 0 & 1 \ep$} (J)
(E) edge node[above=.7em]  {$\bpp z_1 & \varphi \\ 0 & z_2 \epp$} (F)
(I) edge node[auto] {$\cdot z_2$} (J)
(AA) edge[thick, red] (BB)
(BB) edge[thick, blue] (CC)
;
\end{tikzpicture}\vspace{.2cm}
\ee
This diagram is a so-called \emph{distinguished triangle}, which in this case means that it represents a vertically drawn short exact sequence of three sheaves: $\cB_1$, a non-trivial bound state, and $\cB_2$. Hence, elements of Ext$^1(\cB_2, \cB_1)$ are simply all possible entries $\varphi$ in position $(1,2)$ in the tachyon, modulo homotopies. The homotopies amount to adding multiples of rows and columns amongst each other. Here, we see that any dependence of $\varphi$ on $z_1$ or $z_2$ can be washed away via a homotopy. This means that $\varphi$ is a section of $\cO(n_1-n_2-1)$ over the $\P^1$ at the ideal $(z_1, z_2)$. 

Similarly, one can show that ${\rm Ext}^2(\cB_2, \cB_1) \cong {\rm Ext}^1(\cB_1, \cB_2)$ is represented by all possible $(2,1)$ entries in the tachyon modulo the ideal $(z_1, z_2)$.

Now, we would like to compute all of this data directly in F-theory, without making reference to the type IIB information. The F-theory lift can be locally modeled analogously to the one in the previous section, by fibering a conifold over the matter curve. So let us define our CY fourfold as the following hypersurface:
\be 
uv+z_1 z_2 =0
\ee
inside the ambient fivefold
\begin{center}
$X_5: \quad$ \begin{tabular}{c|c|c|c|c|c}
$u$ & $v$ & $\sigma_1$ & $\sigma_2$ & $z_1$ & $z_2$ \\
\hline
$-1$& $-1$ & $1$ & $1$ & $-1$ & $-1$
\end{tabular}
\end{center}
Note that, in this ambient space, the hypersurface we define is CY, since it has homogeneous degree $-2$. A word of caution is in order: Even though a patch of singular  elliptic fibration will usually take this form, the converse may not be true. In this case, it is not possible to complete this non-compact fibration into an elliptic one. This would entail defining $X \equiv u+v$ and $Y \equiv u-v$, and adding an $X^3$ term. However, such a term would have degree $-3$, which is inconsistent. The underlying reason for this obstruction is that, despite being non-compact, this model already has some non-trivial topology from the $\P^1$. The normal bundles of the D7-branes are non-trivial, and this is creating a non-trivial D7-tadpole that must be solved. In other words, in this model, it is not possible to define a consistent axio-dilaton profile around the D7-branes. Nevertheless, since the problems we are interested in can be addressed without canceling D7 tadpoles, we will take this as a toy model for the computation of chiral spectra\footnote{Similarly, we will not discuss D3-brane tadpoles, as they play no relevant role in our analysis, regardless the model being compact or non-compact.}.

Now we wish to lift the matrix factorization \eqref{mfresconifold}, describing the intersecting brane system to a matrix factorization in F-theory. We apply the concrete formula \eqref{upstairs} for Kn\"orrer's periodicity and obtain
\be 
\left( \bpp \phi & 0 \\ 0 & \psi \epp, \bpp \psi & 0 \\ 0 & \phi \epp \right)\,,
\ee
with
\be 
\phi \equiv \bp z_1 & -u \\ v & z_2 \ep \quad {\rm and} \quad \psi \equiv \bp z_2 & u \\ -v & z_1 \ep \,,
\ee
just like in the previous example. The only difference is that, now, the matrices are maps between non-trivial bundles over the ambient space $X_5$:
\be
\begin{tikzcd}[ampersand replacement=\&, column sep =huge]
\begin{matrix} \cO(n_1+1) \\ \oplus \\ \cO(n_1+1)  \end{matrix} \rar{\phi} \& \begin{matrix} \cO(n_1) \\ \oplus \\ \cO(n_1)  \end{matrix}
\qquad 
{\rm and} \qquad 
\begin{matrix} \cO(n_2+1) \\ \oplus \\ \cO(n_2+1)  \end{matrix} \rar{\psi} \& \begin{matrix} \cO(n_2) \\ \oplus \\ \cO(n_2)  \end{matrix}
\end{tikzcd}
\ee
Let us denote the cokernel sheaves of these two complexes by $M$ and $\tilde{M}$, respectively. Then, the full system is specified by the sheaf $M_{\rm tot} = M \oplus \tilde M$. The sheaves $M$ and $\tilde M$ are almost line bundles over the fourfold. They fail to be locally free because their ranks jump from one to two over the singularity. This can be seen as follows: The matrices $\phi$ and $\psi$ generically have rank two over the ambient space, which means there is no cokernel left. Over the hypersurface $uv+z_1 z_2$, they have rank one, leaving a cokernel of rank one. This means they are basically line bundles over the fourfold. At the origin, the matrices vanish, leaving each a cokernel of rank two. 

Just as in the affine case, these matrices tell us about two families of non-Cartier divisors, which in turn correspond to two $U(1)$'s in the effective theory. Following our proposal, the light fields associated to M2's emanating from this singularity should be given by the group Ext$^1(M_{\rm tot}, M_{\rm tot})$  \emph{in the stable category} \underline{MF}. This group decomposes into
\be
{\rm Ext}^1(M_{\rm tot}, M_{\rm tot}) = {\rm Ext}^1(M, M) \oplus {\rm Ext}^1(\tilde M, \tilde M) \oplus {\rm Ext}^1(M, \tilde M) \oplus {\rm Ext}^1(\tilde M, M)\,.
\ee
The chiral and anti-chiral matter given by light M2-branes in which we are interested will sit inside ${\rm Ext}^1(\tilde M, M)$ and ${\rm Ext}^1(M, \tilde M)$, respectively. Hence, we will compute Ext$^1(\tilde{M}, M)$ directly in the fourfold, without making reference to the original type IIB system. The chiral modes in Ext$^1(\tilde{M}, M)$ are given by coherent states of M2-branes, which, as discussed in the introduction, are the lifts of the bifundamental strings trapped at the matter curve. After a tedious calculation, one can show that any element of Ext$^1(\tilde{M}, M)$ is represented by a map of the form

\be \label{LiftChiralGluing}
\begin{tikzpicture}[baseline=(current bounding box.center), row sep=3em, column sep=.5em, text height=1.5ex, text depth=0.25ex]
\node (A) at (0,0) {$\cO(n_1+1)$};
\node (B) at (0,-1)  {$\cO(n_1+1)$};
\node (C) at (0,-2)  {$\cO(n_2+1)$};
\node (D) at (0,-3)  {$\cO(n_2+1)$};
\node (Aa) at (6,0) {$\cO(n_1)$};
\node (Bb) at (6,-1)  {$\cO(n_1)$};
\node (Cc) at (6,-2)  {$\cO(n_2)$};
\node (Dd) at (6,-3)  {$\cO(n_2)$};
\node (p1) at (0,-.5) {$\oplus$};
\node[minimum size=5em] (p2) at (0,-1.5) {$\oplus$};
\node (p3) at (0,-2.5) {$\oplus$};
\node (pp1) at (6,-.5) {$\oplus$};
\node[minimum size=5em] (pp2) at (6,-1.5) {$\oplus$};
\node (pp3) at (6,-2.5) {$\oplus$};

\path[->,font=\scriptsize]
(p2) edge[->] node[above=1.5em] {$\bp z_1 & -u & \varphi & 0 \\ v & z_2 & 0 & \varphi \\ 0 & 0 & z_2 & u\\ 0 & 0 & -v & z_1   \ep$} (pp2);
\end{tikzpicture}\vspace{.2cm}
\ee
with $\varphi \in \cO(n_1-n_2-1)$. To work this out explicitly, one has to use the fact that, in the stable category, \eqref{StableCatEquiv} holds, which in this case means
\be\label{Ext=Hom}
{\rm Ext}^1(\tilde{M}, M)\equiv {\rm Hom}(\tilde{M}, M[1]) \cong {\rm Hom}(\tilde{M}, \tilde{M})\,.
\ee
The homotopies, moreover, eliminate any dependence of $\varphi$ on the ideal $(z_1, z_2, u, v)$, thereby localizing $\varphi$ to the $\P^1$. So, in the end, we find 
\be
{\rm Ext}^1(\tilde{M}, M)  \quad \cong \quad H^0(\P^1, \cO(n_1-n_2-1)) \quad \cong \quad \begin{cases} \C^{n_1-n_2} \quad {\rm for} \quad n_1> n_2 \\
0 \quad {\rm for} \quad n_1 \leq n_2 \end{cases}
\ee
which matches perfectly with the type IIB result in \eqref{iibext}. Similarly, one finds a matching result for the corresponding anti-chiral fields in Ext$^1(M,\tilde{M})$.
Note that the off-diagonal block of the matrix in \eqref{LiftChiralGluing} has exactly the form expected from Kn\"orrer's periodicity for morphisms, i.e. formula \eqref{KnoerrerMorphisms}. Indeed, by the equivalence \eqref{StableCatEquiv}, computing Ext$^1(\tilde{M}, M)$ just means lifting the morphism $(\varphi,\varphi):(z_2,z_1)\to(z_2,z_1)\;\in{\rm Hom}(\mathcal{B}_2,\mathcal{B}_2)$.

Let us pause to summarize the proposal. Given a singular F-theory fourfold, one can construct a catalogue of all possible irreducible matrix factorizations up to isomorphism. These MF's, correspond to sheaves over the singular manifold, that are linked by maps classified by Ext$^*$ groups. Our proposal is that an F-theory compactification is not only a choice of a fourfold plus $C_3$-form. One must supplement this data with \emph{a choice of matrix factorization of the fourfold} that tells us about coherent states of vanishing M2-branes.

\subsection{A class of compact models}\label{U1restrictionSection}
The previous sections covered a prototype F-theory model for two intersecting D7-branes in a non-compact base. The perfect match between the F-theory calculation and the IIB expectation was guaranteed to us by Kn\"orrer's periodicity. The fact that the hypersurface equations describing the D7-branes and the F-theory fourfold were related by adding a $u\,v$ term is very special, and can only be accomplished locally. 

In this section, we will present a class of \emph{globally} defined F-theory fourfolds carrying a family of conifold singularities. They are referred to in the literature as `$U(1)$-restricted' models \cite{Grimm:2010ez}. We will compute the chiral spectrum of such models directly in the singular F-theory background, using the language of matrix factorizations. We will find that this spectrum is localized precisely along the curve of singularities, just as expected. 

In order to check our results, we will proceed to study Sen's weak coupling limit \cite{Sen:1996vd} of these models, which turns out to describe D7/orientifold image D7 pairs. We will discover that the chiral spectrum computed via matrix factorizations in F-theory matches perfectly with the one expected in the perturbative situation.

Let us begin by defining our generic smooth F-theory fourfold as a hypersurface given by
\be
Y^2 = X^3+a_2\, X^2\,Z^2+ a_4\,X\,Z^4 + (a_6+a_3^2)\,Z^6\,.
\ee
where the ambient space is a $\P^2_{2,3,1}$-bundle over some K\"ahler threefold $B_3$. The coordinates $X, Y, Z$ parametrize the projective fiber, and $Z$ transforms as a section of the canonical bundle $K_{B_3}$ of the base. The $a_i$ are sections of $K_{B_3}^{-i}$. They do not correspond to the usual basis of Tate coefficients in the literature, and moreover the $Y$ coordinate has been shifted w.r.t. the one of the Tate polynomial.

The so-called $U(1)$-restriction of \cite{Grimm:2010ez} corresponds to setting $a_6 \equiv 0$. This makes the fourfold singular, with a family of conifold singularities over a curve of $B_3$. This is best seen by rewriting the hypersurface as follows:
\be\label{U1restrTateModel}
(Y+a_3\,Z^3)\,(Y-a_3\,Z^3) = X\,(X^2+a_2\, X\,Z^2+ a_4\,Z^4)\,.
\ee
This has the characteristic $AB = CD$ form of the conifold. For convenience, let us define the following polynomials
\be
Y_\pm \equiv Y \pm a_3\,z^3 \, \quad {\rm and} \quad Q \equiv X^2+a_2\, X\,Z^2+ a_4\,Z^4\,.
\ee
Now we see that this fourfold has two basic matrix factorizations $(\phi, \psi)$ and $(\psi, \phi)$, with
\be
\phi = \bpp Y_+ & Q \\ X & Y_- \epp\, \qquad {\rm and} \qquad \psi = \bp Y_- & -Q \\ -X & Y_+ \ep\,,
\ee
whose associated cokernel sheaves we call $M,\tilde{M}$ respectively. In order to fully specify an MF, we must also fix the domain and codomain of the matrices, which will be vector bundles over the ambient space of the CY fourfold. Let us choose these as follows: Let $\cL$ be an input line bundle that is part of the choice of MF, $H$ the hyperplane bundle of the fiber $\P^2_{2,3,1}$, and $\cO$ the trivial line bundle. Then, we regard the matrices $\phi$ and $\psi$ as the following bundle maps:
\be
\begin{tikzcd}[column sep =40pt, row sep=30pt] 
\cL^{-1} \otimes {K_{B_3}}^{-2} \otimes E \rar{\psi} & \cL^{-1} \otimes {K_{B_3}}^{-2} \otimes  F\,, \\
\cL \otimes {K_{B_3}}^2 \otimes  E \rar{\phi} & \cL \otimes {K_{B_3}}^2 \otimes  F
\end{tikzcd}
\ee
where 
\be\label{EFbundles}
E =  H^{-2} \oplus H^{-3}  \qquad {\rm and} \qquad F =  H \oplus \cO\,.
\ee
Physically, this extra data is fixing a choice of $G_4$ flux in our F-theory background. In \cite{Braun:2011zm, Braun:2014pva}, it was found that one can construct vector bundles $V$ over the CY fourfold via matrix factorizations, such that their second Chern class gives the flux, i.e. $G_4 \equiv c_2(V)$. In this case, the full matrix factorization for this background is the direct sum $(\phi, \psi) \oplus (\psi, \phi)$, with domain and codomain equal to the sum of the respective bundles. Said differently, we specify a background by placing the sheaf 
\be M_{\rm tot} = M \oplus \tilde M \ee on the fourfold.
Following our proposal, the light fields associated to M2's emanating from this singularity should be given by the group Ext$^1(M_{\rm tot}, M_{\rm tot})$ \emph{in the stable category} \underline{MF}. This group decomposes into
\be
{\rm Ext}^1(M_{\rm tot}, M_{\rm tot}) = {\rm Ext}^1(M, M) \oplus {\rm Ext}^1(\tilde M, \tilde M) \oplus {\rm Ext}^1(M, \tilde M) \oplus {\rm Ext}^1(\tilde M, M)\,.
\ee
The chiral and anti-chiral matter given by light M2-branes, in which we are interested, will sit inside ${\rm Ext}^1(\tilde M, M)$ and ${\rm Ext}^1(M, \tilde M)$, respectively. Concretely, the chiral matter is given by all possible vertical maps $\varphi$ modulo homotopy in the following diagram
\be
\begin{tikzcd}[column sep =40pt, row sep=30pt] 
&  \cL^{-1} \otimes {K_{B_3}}^{-2} \otimes E \dar{\varphi} \rar{\psi} & \cL^{-1} \otimes {K_{B_3}}^{-2} \otimes  F \\
\cL \otimes {K_{B_3}}^2 \otimes  E \rar{\phi} & \cL \otimes {K_{B_3}}^2 \otimes  F &
\end{tikzcd}
\ee
By the same type of calculation that led us to \eqref{LiftChiralGluing}, which uses the isomorphism \eqref{Ext=Hom} and the homotopy equivalences, we find:
\be\label{chiralMode}
\varphi  = \rho \cdot \1_2 \quad {\rm with} \quad \rho \in H^0(\mathcal{C}\,,\, \cL^2 \otimes {K_{B_3}}^4 \otimes H^3) \cong H^0(\mathcal{C}\,,\, \cL^2 \otimes {K_{B_3}})\,,
\ee
where $\mathcal{C}$ is the curve given by the ideal $(Y_+, Y_-, X, Q) \cong (Y, X, a_3, a_4)$. In the last equality above we used the fact that, on the curve $\mathcal{C}$, the coordinate $Z$ cannot vanish, and hence $H|_{\mathcal{C}}\cong {K_{B_3}}^{-1}|_{\mathcal{C}}$.  
Similarly, by computing Ext$^1(M,\tilde{M})$, we find anti-chiral fields localized on the same curve:
\be\label{anti-chiralMode}
\tilde \varphi  = \tau \cdot \1_2 \quad {\rm with} \quad \tau \in H^0(\mathcal{C}\,,\, \cL^{-2} \otimes {K_{B_3}}^{-7}) \cong H^1(\mathcal{C}\,,\, \cL^2 \otimes {K_{B_3}})\,,
\ee
where, in the last equality, we used the fact that $K_{\mathcal{C}}\cong {K_{B_3}}^{-6}|_{\mathcal{C}}$. We would like to stress that the localization of these modes $\varphi$ and $\tilde \varphi$ to the curve $\mathcal{C}$ comes out from the calculation of the Ext's, and is \emph{not} enforced by hand. The same remark holds for the calculations in \ref{ConifoldAffine} and \ref{4dchirality}.

Another way to represent these degrees of freedom is as off-diagonal entries in the following matrix
\be
\bpp \phi & \varphi\in{\rm Ext}^1(\tilde{M},M) \\ \tilde{\varphi}\in{\rm Ext}^1(M,\tilde{M}) & \psi \epp\,.
\ee
We can readily compute the chiral index, i.e. the net number of chiral versus anti-chiral fields, using the Riemann-Roch theorem:
\be\label{IndexU1}
{\rm dim\,Ext}^1(\tilde{M},M)-{\rm dim\,Ext}^1(M,\tilde{M})=2\int_\mathcal{C}c_1(\cL)-2c_1(B_3)\,.
\ee
In order to interpret these results, let us take Sen's weak coupling limit of the class of F-theory models in question. This is type IIB string theory on a CY threefold with an O7$^{-}$-plane and a pair of D7/image D7-branes. The CY threefold is a double cover of $B_3$, obtained by introducing one homogeneous coordinate, $\xi$, and one equation, $\xi^2=4a_2$. The O7-plane wraps the locus $\{\xi=0\}$, which lies in the pull-back of the anti-canonical class of $B_3$. The D7 pair, carrying a $U(1)$ gauge theory, is described by the following complex \cite{Collinucci:2008pf,Braun:2011zm}
\be\label{TachyonU1restr}
\begin{tikzcd}[ampersand replacement=\&, column sep =50pt]
\begin{matrix} \cL^{-1} \\ \oplus \\{K_{B_3}}^{4}\otimes \cL  \end{matrix} \rar{T} \& \begin{matrix} \cL \\ \oplus \\ {K_{B_3}}^{-4}\otimes\cL^{-1}  \end{matrix}
\end{tikzcd}
\ee
where the tachyon map is
\be\label{TachyonU1restrMatr}
T=\bpp 0 & -a_4+\xi a_3 \\ a_4+\xi a_3 &0 \epp\,.
\ee
Hence the D7-brane and its orientifold image wrap the loci $S_{\pm}:\{a_4=\pm\xi a_3\}$. They intersect both on the O7-plane and outside it, but it is only the intersection locus outside the O7-plane which carries matter charged under the $U(1)$. This locus is the curve of $B_3$ given by the ideal $(a_3,a_4)$, which coincides with the curve $\mathcal{C}$ we found above in the F-theory analysis. In order to find the matter particles localized on this curve, we can compute Ext$^1({\rm coker}\,T,{\rm coker}\,T)$ and focus on the diagonal fluctuations of the tachyon \eqref{TachyonU1restrMatr}. Taking into account the orientifold invariance\footnote{The wavefunction of fields from open strings stretching between a single D7-brane and its image must vanish on the O7-plane.}, the result is
\be
\delta T=\xi\bpp \rho & 0 \\ 0 &\tau \epp\,,
\ee
where $\rho\in H^0(\mathcal{C}\,,\, \cL^2 \otimes {K_{B_3}})$ and $\tau\in H^0(\mathcal{C}\,,\, \cL^{-2} \otimes {K_{B_3}}^{-7})$. This perfectly matches the results of our MF computation in F-theory, i.e. \eqref{chiralMode} and \eqref{anti-chiralMode}. Finally, using \eqref{TachyonU1restr}, we find for the net D7-brane gauge flux
\be
F_{S_+}-F_{S_-}=2F_{S_+}=2c_1(\cL)-4c_1(B_3)\,,
\ee
which generates the same index we computed in \eqref{IndexU1}.

\subsection{Global T-branes and point-like matter}\label{GlobalTbraneSec}

In this section we would like to show the power of MF's to investigate F-theory backgrounds, which are inaccessible if we resolve or deform the fourfold. We will do that in the compact setting of section \ref{U1restrictionSection}, i.e. without relying on Higgs profiles of any (local) gauge theory perspective.

Using the massless fields we found in section \ref{U1restrictionSection}, we can engineer T-brane backgrounds in this global context. For example, we may give a non-trivial vev just to $\rho$, and thus define a compact F-theory background by the following MF
\be\label{MFGlobalPointLike}
\left(\bpp \phi & \rho\1_2 \\ 0 & \psi \epp\, , \, \bpp \psi & -\rho\1_2 \\ 0 & \phi \epp\right)\,.
\ee
This background breaks the $U(1)$ gauge group and obstructs the small resolutions of the fourfold \eqref{U1restrTateModel}. We now want to study the spectrum of fluctuations around it. We will discover a very peculiar phenomenon, i.e. the existence of matter trapped at points in $B_3$, which defies the standard F-theory paradigm of matter being localized on curves. In the local context of 7-brane gauge theories, T-brane backgrounds have already been shown to feature this kind of behavior \cite{Cecotti:2010bp}. In what follows we will provide the first instance of this in a globally defined F-theory model, and give a conjecture on its physical origin.

Let $\mathcal{F}$ be the cokernel sheaf associated to the MF\footnote{The cokernel sheaf associated to an MF is by definition the cokernel sheaf of the first matrix in the factorization.} \eqref{MFGlobalPointLike}. General fluctuations around the background specified by $\mathcal{F}$ correspond to elements of the group Ext$^1(\mathcal{F},\mathcal{F})$ (in the \emph{full} category of MF's). In cases where there is a non-trivial unbroken gauge group, this would be the adjoint matter spectrum of the system. Finding this spectrum means counting all possible vertical maps $\Delta$ modulo homotopy in the following diagram
\be\label{Ext1FF}
\begin{tikzcd}[ampersand replacement=\&,column sep =50pt, row sep=40pt] 
\&  \mathcal{V} \dar{\Delta} \rar{\bpp \phi & \rho\1_2 \\ 0 & \psi \epp} \& \mathcal{W} \\
\mathcal{V} \rar{\bpp \phi & \rho\1_2 \\ 0 & \psi \epp} \& \mathcal{W} \&
\end{tikzcd}
\ee
where $\mathcal{V}$ and $\mathcal{W}$ are the following rank-four vector bundles:
\be
\mathcal{V} =  \cL\otimes {K_{B_3}}^{2} \otimes E \,\oplus \, \cL^{-1}\otimes {K_{B_3}}^{-2} \otimes F\,, \qquad \mathcal{W} =   \cL\otimes {K_{B_3}}^{2} \otimes F \,\oplus\,  \cL^{-1}\otimes {K_{B_3}}^{-2} \otimes E\,,
\ee
with $E,F$ as in \eqref{EFbundles}. As already stressed at the beginning of section \ref{ConifoldSection}, these fluctuations may be distributed in two classes: The ones which deform the polynomial defining the fourfold, and the ones which do not. The former are associated to complex structure moduli of the F-theory geometry, whereas the latter are given by light M2-branes stuck at the singularities, missed by the supergravity analysis. We argue that the latter class of fluctuations, which is our main interest in this paper, are captured by elements of Ext$^1(\mathcal{F},\mathcal{F})$ that survive the quotient MF$\to$\underline{MF} (defined in section \ref{KnoerrerSection}) of the fourfold polynomial. Therefore, we will now pass to the \emph{stable category} \underline{MF}.

In order to compute these M2 degrees of freedom, we proceed by exploiting Kn\"orrer's periodicity. Indeed, since the fourfold \eqref{U1restrTateModel} has the characteristic form of the conifold, we can still reduce our problem to a lower dimensional one, even though now, in contrast to sections \ref{ConifoldAffine} and \ref{4dchirality}, this procedure has in principle nothing to do with going to the weakly coupled type IIB description. With this method, we will discover point-like matter in this system.

First of all, we observe that, due to \eqref{StableCatEquiv}, $\Delta$ is part of the following morphism of MF's
\be
(\tilde{\Delta},\Delta):\quad\left(\bpp \phi & \rho\1_2 \\ 0 & \psi \epp\, , \, \bpp \psi & -\rho\1_2 \\ 0 & \phi \epp\right)\longrightarrow \left(\bpp \psi &- \rho\1_2 \\ 0 & \phi \epp\, , \, \bpp \phi & \rho\1_2 \\ 0 & \psi \epp\right)\,,
\ee
where $\tilde{\Delta}$ is the `partner' morphism in the sense of \eqref{MFmorphisms}. Let us rearrange rows and columns of our starting MF \eqref{MFGlobalPointLike} to write it as follows:
\be\label{MFGlobalPointLikeNew}
\left(\left(\begin{array}{cc}\begin{array}{cc}Y_+&\rho\\0& Y_- \end{array}&Q\,\1_2\\ X \, \1_2 & \begin{array}{cc}Y_-&-\rho\\0& Y_+ \end{array}\end{array}\right)\;,\;\left(\begin{array}{cc}\begin{array}{cc}Y_-&-\rho\\0& Y_+ \end{array}&-Q\,\1_2\\ -X \, \1_2 & \begin{array}{cc}Y_+&\rho\\0& Y_- \end{array}\end{array}\right)\right)\,.
\ee
Now we immediately realize that this MF has the same form as in \eqref{upstairs}, with $Q$ and $X$ playing the role of $-u$ and $v$ respectively. Therefore, calling $(\tilde{\Delta'},\Delta')$ the pair of maps $(\tilde{\Delta},\Delta)$ in the new basis, formula \eqref{KnoerrerMorphisms} tells us that
\be
(\tilde{\Delta'},\Delta')=\left( \bpp \tilde{\delta} & 0 \\ 0& -\delta \epp\,, \bpp \delta & 0 \\ 0  & - \tilde{\delta} \epp \right)\,,
\ee
where the pair $(\tilde{\delta},\delta)$ is the following morphism of MF's of the polynomial $Y_+\cdot Y_-$:
\be\label{IntermediateStep}
\begin{tikzcd}[ampersand replacement=\&,column sep =55pt, row sep=45pt]
G \otimes H^{-2} \rar{\bpp Y_+ & \rho \\ 0 & Y_- \epp} \dar{\delta} \& G \otimes H \dar{\tilde{\delta}}\\ G \otimes H \rar{\bpp Y_- & -\rho \\ 0 & Y_+ \epp} \& G \otimes H^{4}
\end{tikzcd}
\qquad {\rm with}\qquad G= \left[\begin{array}{ccc}\cL&\otimes&{K_{B_3}}^{2}\\ &\oplus&\\ \cL^{-1}&\otimes&{K_{B_3}}^{-2} \end{array}\right]\,.
\ee
So far we have managed to get rid of the variables $X$ and $Q$, since Kn\"orrer's periodicity guarantees that all maps in \eqref{IntermediateStep} are independent of them. But actually we can do better. Let us use Kn\"orrer's periodicity once more, this time with $Y_+$ and $Y_-$ playing the role of $u$ and $v$. Again, formula \eqref{KnoerrerMorphisms} tells us that\footnote{The reason why $c$ and $\tilde{c}$ are off-diagonal, as opposed to diagonal, is because the variables we are eliminating by Kn\"orrer's periodicity this time are diagonal, as opposed to off-diagonal.}
\be
(\tilde{\delta},\delta)=\left( \bp 0 & c \\ -\tilde{c}& 0 \ep\,, \bp 0 & c \\ -\tilde{c}  & 0 \ep \right)\,,
\ee
where the pair $(\tilde{c},c)$ is the following morphism of MF's of the \emph{zero} polynomial:
\be\label{FinalStep}
\begin{tikzcd}[column sep =40pt, row sep=35pt]
 \cL^{-1}\otimes{K_{B_3}}^{-2}\otimes H^{-2} \rar{\rho}  \dar{c} & \cL\otimes{K_{B_3}}^{2} \otimes H \dar{\tilde{c}} \dlar[dashed, near start] \\ \cL\otimes{K_{B_3}}^{2} \otimes H \rar{0} & \cL^{-1}\otimes{K_{B_3}}^{-2} \otimes H^{4}
\end{tikzcd}
\ee
The problem is now reduced to the much easier one of determining the pair of maps $(\tilde{c},c)$ in \eqref{FinalStep} up to homotopy. Kn\"orrer's periodicity already guarantees that $\tilde{c}$ and $c$ are independent of $Y_+,Y_-,X,Q$. Moreover, commutativity of \eqref{FinalStep} clearly implies $\tilde{c}=0$, while homotopy eliminates any dependence of $c$ on $\rho$.

Hence, our final result for the map $\Delta$, in the original basis of \eqref{Ext1FF}, is
\be\label{PointLikeMatterGlobal}
\Delta=\bpp 0&c\,\1_2 \\ 0 & 0\epp\,,\qquad c\in H^0(\mathcal{P},\cL^{2}\otimes{K_{B_3}}^{4}\otimes H^{3})\equiv \C^{p}\,, 
\ee
where $\mathcal{P}$ is the set of points in $B_3$ given by the ideal $(Y_+,Y_-,a_3,a_4,\rho)$, and $p$ is the number of such points. In other words, we have found that, in the stable category \underline{MF}
\be\label{ExtPointLike}
{\rm Ext}^1(\mathcal{F},\mathcal{F})=\C^p\,.
\ee
As promised, with our technique we have discovered matter modes concentrated at points of the F-theory `internal' space. In \cite{Collinucci:2014qfa}, we gave a physical explanation, alternative to the one based on the Higgs profile \cite{Cecotti:2010bp}, of the analogous phenomenon taking place in T-brane backgrounds of 7-brane gauge theories. There we argued that, actually, in these situations, there are anti-D3-branes\footnote{Their nature of anti-D3's, as opposed to D3's, came from analyzing the positions of the complexes defining them in the derived category of coherent sheaves: It turned out that the 3-branes which are mutually supersymmetric with the considered system of intersecting D7-branes are anti-D3-branes (and not D3-branes).} located at those points, and the point-like fluctuations which arise, are nothing but the degrees of freedom of movement of the anti-D3's along a curve of the internal space. Since we do not have any Higgs field to invoke in this case, our argument of embedded lower dimensional branes seems to be the only available one to explain the appearance of the point-like matter \eqref{ExtPointLike}.
By M/F-theory duality, indeed, we deduce that the fluctuations $c$ of \eqref{PointLikeMatterGlobal} are due to the degrees of freedom of anti-M2-branes located at $\mathcal{P}$ to move along the curve $\mathcal{C}$.

\section{Matrix factorizations vs resolutions}\label{ResolutionSection}

In this section we present further evidence for the appropriateness of the matrix factorization machinery in dealing with F-theory singularities. We will show how the structures we add to the singular space are not \emph{ad hoc}, but rather they are deeply and beautifully related to the geometry of its resolution. The most fundamental framework for connecting a singular variety to its resolution in this context is the language of \emph{non-commutative crepant resolutions} \cite{Bergh:aa}. For the sake of simplicity, we will refrain from introducing the whole formalism, but only use it implicilty. 

We will work in affine space with a very familiar class of singularities, over which we have complete control at weak coupling: The singularities belonging to the {\bf A} series of the {\bf ADE} classification. The low-energy theory corresponding to the {\bf A$_{\bf n-1}$} singularity is a $U(n)$ gauge theory\footnote{The center of mass $U(1)$ in $U(n)$ decouples, leaving an interacting $SU(n)$ theory.} dual to the worldvolume theory of a stack of $n$ D7-branes. As in section \ref{ConifoldAffine}, we will start with type IIB and then uplift to F-theory using Kn\"orrer's periodicity.

Consider type IIB on $\mathbb{R}^{1,7}\times\mathbb{C}$; with $n$ coincident D7-branes at the origin of $\C$, and discard the irrelevant longitudinal $\mathbb{R}^{1,7}$ factor. Let $S\equiv \mathbb{C}[z]$ be the ring of functions in one complex variable along the transverse $\C$. The zero-dimensional space describing the full system of branes is the \emph{non-reduced scheme} $z^n=0$. There exist $n-1$ irreducible, inequivalent, non-trivial, reduced MF's of the polynomial $z^n$ given by the pairs $(z^i,z^{n-i})_{i=1,\ldots,n-1}$. Each corresponds to `picking-up' a sub-stack made of $i$ out of the $n$ branes of the original stack. Their associated cokernel sheaves, which we will call $M_i$ in the following, are just the coordinate rings of such sub-stacks. 
\be
(z^i, z^{n-i}) \quad \in \quad {\rm MF}(z^n) \qquad \longleftrightarrow \qquad M_i \equiv {\rm coker}(z^i)\,.
\ee
As usual, there are also the two uninteresting MF's, which give no information about the structure of the space in question: $(1,z^n)$ corresponding to the empty brane, and $(z^n,1)$ associated to the coordinate ring $M_n= R \equiv S/(z^n)$ of the entire stack.

The D7-brane system giving rise to the $U(n)$ gauge theory results from the following tachyon condensation process
\be
\begin{tikzcd}
0 \arrow{r} & S^{\oplus n} \arrow{r}{T} & 
S^{\oplus n} \arrow{r} & \mathcal{S}_{U(n)} \rar & 0\,,
\end{tikzcd}
\ee
where $T$ is the tachyon with profile 
\be T = z   \cdot \1_n \,.\label{suntachyon}\ee
Each diagonal entry in \eqref{suntachyon} should be regarded as the tachyon for each D7-brane of the stack taken individually. To see this more clearly, we can go to the Coulomb branch of the theory, where the $n$ D7-branes are displaced over $n$ distinct points $\{z_k\}_{k=1,\ldots,n}$ of $\mathbb{C}$. In this branch, the tachyon becomes
\be \label{coulombtachyon}
T = {\rm diag}(z-z_1, \ldots, z-z_n)\,.
\ee
Regarding the tachyon as the first matrix of the pair defining an MF (see section \ref{RelationToD7s}), equation \eqref{coulombtachyon} corresponds to the reducible MF given by the direct sum  
\be
\bigoplus_{k=1}^n \biggl(\,z-z_k\;, \, \quad (z-z_1)\cdots (z-z_{k-1})(z-z_{k+1})\cdots (z-z_n)\,\biggr)\,.
\ee
The sheaf associated to the $k$th summand is the coordinate ring of the $k$th brane of the stack, which will be named $L_k$: 
\be
L_k\,: {\rm coker}(z-z_k)\,.
\ee
The tachyon of the entire system thus treats all components of the stack \emph{democratically}. It is easy to see that such $L_k$ can be represented as the following cokernel sheaves
\be\label{Lk}
\begin{tikzcd}
0 \arrow{r} & M_{k-1} \arrow{r} & 
M_{k} \arrow{r} & L_k \rar & 0\,,
\end{tikzcd}
\ee
where $M_0$ is defined to be the empty set. Of course, in the $U(n)$ phase, since all branes become indistinguishable, the $L_k$ all look like copies of $M_1$. In other words, if all $z_k$ are equal, the cokernels of the maps $M_{k-1}\to M_k$ are all isomorphic to $M_1$. 

Let us now uplift this information to F-theory. In complete analogy to the conifold example of section \ref{ConifoldAffine}, we focus on the vicinity of the singular fibers and write the elliptic fibration as the following subspace of $\mathbb{C}^3$
\be\label{AK3}
P_{K3} \equiv uv-\prod_{i=1}^{n}(z-z_i)=0\,,
\ee
that represents a local K3 surface, which is a limit of the $n$-centered Taub-NUT space TN$_{\bf n}$. In the limit where all $z_i$ become equal, we get an ${\bf A}_{\bf n-1}$ singularity at the origin. Note, that a K3 surface can also have singularities of type {\bf D} and {\bf E}, which give rise to the corresponding gauge group enhancements via M2-branes wrapped on the vanishing cycles. However, these geometries do not admit descriptions as multi-centered Taub-NUT spaces. In particular, the {\bf E} case does not even admit a circle fibration, in accordance with the lack of a perturbative D-brane description of such gauge groups.

Now we will uplift the matrix factorization data describing the tachyon, and hence the D7-branes, onto matrix factorizations of the K3 surface. In order to do so, we will apply   formula \eqref{upstairs}.

To avoid cluttering, we keep the same name for the lifts of the various objects. Therefore $S=\mathbb{C}[u,v,z]$ is the ring of functions in three complex variables, $R=S/(P_{K3})$ is the structure sheaf of the local K3, and the $\{M_i\}_{i=1,\ldots,n-1}$ are a basis of sheaves associated to all the irreducible MF's of the polynomial in \eqref{AK3}, i.e.
\begin{eqnarray}\label{MFA}
M_i&=&{\rm coker} \,(\phi_i\,,\,\psi_i)\,, \quad{\rm with} \\
  \quad \phi_i&=& \bpp u & \prod_{k=1}^{i}(z- z_i) \\ \prod_{j=i+1}^{n}(z- z_j) & v \epp\,\quad \;\psi_i = \bpp v & -\prod_{k=1}^{i}(z- z_i) \\ -\prod_{j=i+1}^{n}(z- z_j) & u \epp\,. \nonumber
\end{eqnarray}
Note that, as long as the space is non-singular, these MF's can all be transformed to trivial and non-reduced ones. By means of suitable basis redefinitions, we can easily realize that both $M_0$ and $M_n$ are copies of $R$. The sequences \eqref{Lk} remain formally identical, but now with the objects redefined as in \eqref{MFA}, and the $L_k$ are the following sheaves
\begin{equation}
\begin{tikzcd}[column sep=large]
L_k\,: & S^{\oplus 2} \rar{(z_k, u)} & S\,.
\end{tikzcd}
\end{equation}

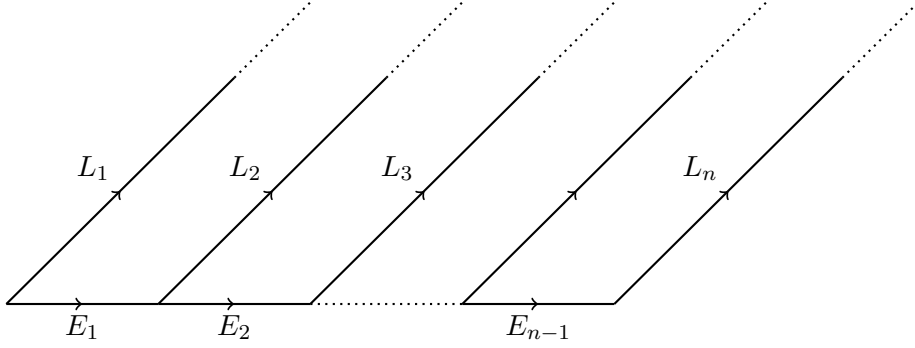
\begin{figure} 
\hspace*{5em}{\begin{tikzpicture}[scale=2]
\draw[->-, thick] (0,0)--(1,0) node[below, midway]{$E_1$};
\draw[->-, thick] (1,0)--(2,0) node[below, midway]{$E_2$};
\draw[-, thick, dotted] (2,0)--(3,0);
\draw[->-, thick] (3,0)--(4,0) node[below, midway]{$E_{n-1}$};

\draw[->-, thick] (0,0)--(1.5,1.5) node[auto, midway]{$L_1$};
\draw[-, dotted, thick] (1.5,1.5)--(2,2);
\draw[->-, thick] (1,0)--(2.5,1.5) node[auto, midway]{$L_2$};
\draw[-, dotted, thick] (2.5,1.5)--(3,2);
\draw[->-, thick] (2,0)--(3.5,1.5) node[auto, midway]{$L_3$};
\draw[-, dotted, thick] (3.5,1.5)--(4,2);
\draw[->-, thick] (3,0)--(4.5,1.5);
\draw[-, dotted, thick] (4.5,1.5)--(5,2);
\draw[->-, thick] (4,0)--(5.5,1.5) node[auto, midway]{$L_n$};
\draw[-, dotted, thick] (5.5,1.5)--(6,2);
\end{tikzpicture}}
\caption{Schematic picture of a multi-centered Taub-NUT space.}
\label{MultiTaubNUTFig}
\end{figure}
Note, that at the moment, we are \emph{not} working in the stable category \underline{MF}$(P_{K3})$. In other words, for the time being, we are \emph{not} modding out copies of $R$.
Each $L_k$ sheaf is the ring of functions of a non-compact two-cycle in the corresponding TN$_{\bf n}$ geometry. Such 2-cycles, which by abuse of notation we also call $L_k$, are obtained for every $k$ by fibering the M-theory circle over non-intersecting lines connecting each D7-brane to infinity \cite{Witten:2009xu}. They are pictorially described in fig. \ref{MultiTaubNUTFig}, where, for the sake of clarity, the D7's have been T-dualized to D6's. 

This geometry also has compact two-cycles $E_k$ with the topology of $S^2$, simply given by fibering the M-theory circle on the lines connecting two branes without meeting any of the others. When we take the limit yielding the ${\bf A}_{\bf n-1}$ singularity, these compact 2-cycles will vanish. If, then, instead of deforming, we resolve the space, we will see these 2-cycles reappear as exceptional divisors. Referring to the orientations displayed in fig. \ref{MultiTaubNUTFig}, one observes that, for $k=1,\ldots,n-1$, the combinations $-L_k+E_k+L_{k+1}$ are trivial in the second homology group of TN$_{\bf n}$. This fact tells us that the coordinate ring of each $E_k$ must be obtained by taking the mapping cone of a non-trivial homomorphism from $L_{k+1}$ to $L_k$. It is very instructive to see in practice how this comes about at the level of complexes.

Let us go to the origin of the Coulomb branch, and focus on the singular K3 surface, given by 
\be \label{singK3}
u v - z^n=0\,.
\ee
Now, the MF's become non-trivial, and are given by the following
\be \label{MFsing}
M_i={\rm cokernel} \,(\phi_i\,,\,\psi_i)\,, \quad\quad \phi_i= \bpp u & z^i \\ z^{n-i} & v \epp\,\quad {\rm and} \quad \;\psi_i = \bpp v & -z^i \\ -z^{n-i} & u \epp\,,
\ee
and all $L_k$ become equivalent:
\bcd[column sep=large]
L_k\,: & S^{\oplus 2} \rar{(z, u)} & S\,.
\ecd
A non-trivial homomorphism from $L_{k+1}$ to $L_k$ is associated to the following commutative diagram
\be
\begin{tikzcd}
S^{\oplus 2} \dar{\mathbb{1}_2} \rar{(z, u)}& S \dar{1}\\
S^{\oplus 2} \rar{(z, u)} & S
\end{tikzcd}
\ee
Using \eqref{Lk} and omitting zeros, we can rewrite this as a commutative diagram between sequences of the $M_i$ as follows:
\be\label{L-L}
\begin{tikzcd}
M_k \dar \rar& M_{k+1} \dar\\
M_{k-1} \rar& M_k
\end{tikzcd}
\ee
Taking the mapping cone of \eqref{L-L} (see section \ref{KnoerrerSection} for the definition) leads us to the following complex
\be\label{AR}
\begin{tikzcd}
M_{k} \arrow{r} & 
M_{k+1}\oplus M_{k-1} \arrow{r} & M_k\,.
\end{tikzcd}
\ee
For $k=1,\ldots,n-1$, \eqref{AR} can be shown to be an \emph{exact sequence} of $R$-modules. Such sequences are objects of fundamental importance in commutative algebra and go under the name of `Auslander-Reiten sequences' (see \cite{MCMbook} for definitions). To see the coordinate ring of $E_k$ emerge from \eqref{AR}, we have to `lift' the sequence to the resolved space. In the lift, the various sheaves $M_k$ become line bundles on the blown-up surface and the sequence ceases to be exact at the middle position. The ensuing cokernel sheaf is supported only over $E_k$. In the following, we will explain the details of this process.

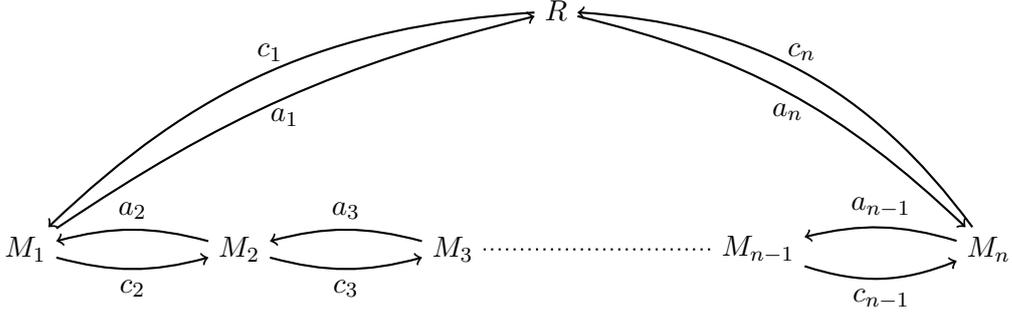
\begin{figure} \hspace*{2em}
\begin{tikzpicture}[scale=5, node distance=2cm,
block/.style={draw, circle,
minimum size=1.1cm,
font=\small}]
  \coordinate (a) at (0,0);
  \coordinate (b) at (3,2);
  \node at (a) (M1) {$M_1$};
  \node[right=of M1] (M2) {$M_2$};
  \node[right=of M2] (M3) {$M_3$};;
    \node[right=3 of M3] (Mn-1) {$M_{n-1}$};;
        \node[right=of Mn-1] (Mn) {$M_n$};;
  \node[above right =2.6 and .67 of M3] (R) {$R$};

 \draw (R) edge[->, bend right=20, thick] node[above]{$c_1$} (M1);
  \draw (M1) edge[->, bend left=10, thick]  node[below]{$a_1$} (R);

 \draw (M1) edge[->, bend right=15, thick] node[below]{$c_2$} (M2);
  \draw (M2) edge[->, bend right=15, thick] node[above]{$a_2$} (M1);
 \draw (M2) edge[->, bend right=15, thick] node[below]{$c_3$} (M3);
  \draw (M3) edge[->, bend right=15, thick] node[above]{$a_3$} (M2);
 \draw (R) edge[->, bend left=15, thick] node[below]{$a_n$} (Mn);
  \draw (Mn) edge[->, bend right=25, thick] node[above]{$c_n$} (R);
 \draw (Mn-1) edge[->, bend right=20, thick] node[below]{$c_{n-1}$} (Mn);
  \draw (Mn) edge[->, bend right=15, thick] node[above]{$a_{n-1}$} (Mn-1);
  \draw (M3) edge[dotted, thick] (Mn-1);
                  
\end{tikzpicture}
\caption{Auslander-Reiten quiver for an A$_{n-1}$ singularity}
\label{fig:AR}
\end{figure}

First we need to know the maps entering \eqref{AR} and second the dictionary translating them into quantities of the resolved space. It is well-known that the Auslander-Reiten sequences can be read-off from the \emph{quiver diagram} associated to a given singularity (see \cite{MCMbook} and \cite{Yoshino}). For singularities of the {\bf ADE} type such quivers have the shape of the extended Dynkin diagram of the corresponding Lie algebra. The nodes are the modules $M_k$, whereas the links are replaced by a pair of oppositely-oriented arrows representing maps between modules. The quiver of the ${\bf A}_{\bf n-1}$ singularity is drawn in figure \ref{fig:AR}, where the extended node is associated to the coordinate ring of the singular space $M_0=M_n=R$, i.e. to the trivial element of \underline{MF}. For $k=1,\ldots,n$, let $c_k:M_{k-1}\to M_k$ \, and \, $a_k:M_{k}\to M_{k-1}$ be the maps between two consecutive modules. There are $n$ relations among these maps, which for each $k$ impose equality of the two shortest loops based on the same module, i.e.\footnote{We adopt the convention that maps on the right act first.}
\be\label{Relations:c-a}
c_k a_k = a_{k+1} c_{k+1}\quad , \quad k=1,\ldots,n\,,
\ee
where the map index is defined modulo $n$. The ${\bf A}_{\bf n-1}$ singularity is deduced from the quiver by looking at the \emph{center} of its path algebra: Such center is indeed generated by three elements subjected to a relation of the form \eqref{AK3}.  Thanks to the relations \eqref{Relations:c-a}, we can re-write the Auslander-Reiten exact sequences \eqref{AR} more explicitly:
\be\label{ARmaps}
\begin{tikzcd}[column sep =huge]
M_{k} \arrow{r}{\bpp c_{k+1}  \\  - a_k \epp} & 
M_{k+1}\oplus M_{k-1} \arrow{r}{(a_{k+1}\,,\,c_k)} & M_k\,,
\end{tikzcd}
\ee
where here $k$ runs from $1$ to $n-1$.\footnote{For $k=n$ the sequence can be shown to be not exact already in the singular space, and to lead, at large volume, to the structure sheaf of the entire exceptional divisor.} The formalism of \emph{non-commutative crepant resolutions} allows us to export sheaves, maps between sheaves, and sequences such as \eqref{ARmaps} to large volume. However, in this particular case, we can take a shortcut around this machinery by working with the toric resolution\footnote{This shortcut is possible because this toric resolution is closely related to the \emph{quiver GIT} construction of the moduli space of representations of the Auslander-Reiten quiver with dimension vector $(1, 1, \ldots, 1)$. See \cite{Wemyss:2012ee} for lecture notes.} of ${\bf A}_{\bf n-1}$, which is defined as the variety given by the following $n-1$ projective relations:
\be\label{ToricBlup}
\begin{array}{ccccccccc}v_0&v_1&v_2&v_3&\cdots&v_{n-3}&v_{n-2}&v_{n-1}&v_n\\ \hline 1&-2&1&0&\cdots&0&0&0&0\\ 0&1&-2&1&\cdots&0&0&0&0\\  \vdots&\vdots&\vdots&\vdots&\ddots&\vdots&\vdots&\vdots&\vdots \\ 0&0&0&0&\cdots&1&-2&1&0\\ 0&0&0&0&\cdots&0&1&-2&1 \end{array}
\ee
in the $n+1$ homogeneous coordinates $v_0,\ldots,v_n$. In terms of them the affine coordinates of \eqref{AK3} are: $u=\Pi_{i=0}^{n}v_i^{n-i}$, $v=\Pi_{i=0}^{n}v_i^{i}$ and $z=\Pi_{i=0}^{n}v_i$. The exceptional divisors $E_k$ of the blow-up are given by the loci $\{v_k=0\}_{k=1,\ldots,n-1}$, while the Stanley-Reisner ideal of the resolved space is made of the following elements
\be\label{SR}
SR\;:\quad\Big\{ v_i\, v_j \Big\}_{j\ge i+2}\;.
\ee
For $k=1,\ldots,n-1$, the sheaf $M_k$ becomes locally free in the resolution and corresponds to the line bundle $\cO_k(1)$, where $\cO_k(d)\equiv\cO(0,\ldots,d,\ldots,0)$, with the degree $d$ in the $k$th position. Its transition functions are given by the $\C^*$ action of the $k$th row of table \eqref{ToricBlup}. The structure sheaf $R$, instead, is sent to the trivial line bundle $\cO$. As for the maps, we assign to them the following `multi-charge': $-1$ w.r.t. the sheaf the map originates from, $1$ w.r.t. the sheaf it ends on, and $0$ w.r.t. all other sheaves. For instance, $c_2$ becomes a section of $\cO(-1,1,\ldots,0)$.
Comparing such degrees with table \eqref{ToricBlup}, one can immediately write down the following dictionary between quiver maps and homogeneous coordinates:
\be\label{Maps-Coordinates}
c_k=\prod_{j=0}^{k-1} v_j \quad {\rm and}\quad a_{k}=\prod_{j=k}^n v_{j},\quad\, k=1,\ldots,n\,.
\ee
We are now finally able to lift the exact sequences \eqref{ARmaps} to the resolved space. Using \eqref{Maps-Coordinates}, we obtain
\be\label{ARBlup}
\begin{tikzcd}[column sep =3 cm]
\cO_{k}(1) \arrow{r}{v_k\bpp v_0\cdots v_{k-1}  \\  - v_{k+1}\cdots v_n \epp} & 
\cO_{k+1}(1)\oplus \cO_{k-1}(1) \arrow{r}{(v_{k+1}\cdots v_n \,, \,v_0\cdots v_{k-1})} & \cO_k(1)\,.
\end{tikzcd}
\ee
The rightmost map in \eqref{ARBlup} is surjective $\forall k=1,\ldots,n-1$, because the two components cannot vanish simultaneously, due to \eqref{SR}. However, exactness is clearly lost in the middle position wherever $v_k=0$. Therefore, we expect this sequence to give us the cokernel sheaf of the map $v_k$, shifted one position to the left. This sheaf is by definition supported only over the locus $E_k:\{v_k=0\}$, which is the $k$th exceptional $\P^1$ of the resolution, as anticipated. To see why this is the case, it suffices to note that the complex \eqref{ARBlup} can be rewritten as the following cone:

\be\label{ConeAR}
\begin{tikzpicture}[baseline=(current bounding box.center), row sep=3em, column sep=2.8em, text height=1.5ex, text depth=0.25ex]
\node (A) at (0,0) {$\cO_k(-1)\otimes\cO_{k+1}(1)$};
\node (B) at (5.75,0) {$\cO_{k+1}(1)\oplus \cO_{k-1}(1)$};
\node (C) at (11.5,0) {$\cO_k(1)$};
\node (p1) at (-.06,-.7) {$\oplus$};
\node (p3) at (5.75,-.7) {$\oplus$};
\node (D) at (0,-1.4) {$\cO_k(1)$};
\node (E) at (5.87,-1.4) {$\cO_k(-1)\otimes\cO_{k+1}(1)$\,.};
\path[->,font=\scriptsize]
(A) edge node[auto] {$\begin{array}{c}\bpp v_0\cdots v_{k-1}  \\  - v_{k+1}\cdots v_n \epp\\ {} \end{array}$} (B)
(B) edge node[auto] {$(v_{k+1}\cdots v_n \,, \,v_0\cdots v_{k-1})$} (C)
(D) edge node[above] {$v_k$} (E)
(A) edge node[above] {$1$} (E)
;
\end{tikzpicture}\vspace{.2cm}
\ee
The upper sequence in \eqref{ConeAR} is now exact at all positions and can be removed, since its cokernel sheaf is empty. We are thus left with the lower complex, which is what we aimed for.

Analogous arguments, though with more involved algebra, can be repeated not only for singularities of type {\bf D} and {\bf E}, but also for non-quotient singularities. 

The conifold is the easiest example of this class, and it is worth sketching its features. We repeat some definitions for convenience:
\be \label{reconifold}
R=\C[u,v,x,y]/(xy+uv)\,, \qquad \phi \equiv \bp x & -u \\ v & y \ep, \quad {\rm and} \quad \psi \equiv \bp y & u \\ -v & x \ep \,.
\ee
For the conifold, aside from the structure sheaf of the singular space, there are two modules, corresponding to the cokernel sheaves of the two irreducible MF's. As in section \ref{4dchirality}, we define $M\equiv{\rm coker}(\phi)$ and $\tilde{M}\equiv{\rm coker}(\psi)$. Again, looking at the quiver diagram associated to this singularity, it is possible to show that there are two `specular' Auslander-Reiten sequences,  which look like
\be\label{AR1}
\begin{tikzcd}
\tilde{M} \arrow{r} & 
R^{\oplus2} \arrow{r} & M\,,
\end{tikzcd}
\ee
and
\be\label{AR2}
\begin{tikzcd}
M \arrow{r} & 
R^{\oplus2} \arrow{r} & \tilde{M}\,,
\end{tikzcd}
\ee
where now $R=\C[u,v,x,y]/(xy+uv)$. Notice that, as opposed to the {\bf A}-type singularities, the Auslander-Reiten sequences for the conifold represent non-trivial extensions of two \emph{different} sheaves, whereas there exists no non-trivial irreducible extension between two copies of the same sheaf. It is interesting to study the fate of \eqref{AR1} and \eqref{AR2} as we go to large volume. Again, the most fundamental framework for doing this is by using non-commutative crepant resolutions. However, for the same reasons as before, by performing a small resolution of the conifold singularity, using the toric language, we will accomplish this promptly. This amounts to introducing the following projective relation
\be\label{ToricBlupC}
\begin{array}{cccc}\Sigma_1&\Sigma_2&\tilde{\Sigma}_1&\tilde{\Sigma}_2 \\ \hline 1&1&-1&-1 \end{array}
\ee
in the four homogeneous coordinates $\Sigma_1,\Sigma_2,\tilde{\Sigma}_1,\tilde{\Sigma}_2$, in terms of which the affine coordinates of \eqref{reconifold} are recovered as: $x=\Sigma_1 \tilde{\Sigma}_1$, $y=\Sigma_2 \tilde{\Sigma}_2$, $u=\Sigma_1 \tilde{\Sigma}_2$, and $v=-\Sigma_2 \tilde{\Sigma}_1$. As is well-known, this smooth geometry has two phases, which we label by a sign, corresponding to the choice of Stanley-Reisner ideal of \eqref{ToricBlupC}: The positive phase where the locus $\{\Sigma_1=\Sigma_2=0\}$ is removed, and the negative phase where instead the locus $\{\tilde{\Sigma}_1=\tilde{\Sigma}_2=0\}$ is removed. The sheaves $M,\tilde{M}$ both become locally free at large volume and are identified with the line bundles $\cO(1),\cO(-1)$ respectively. Examining the charges of the maps involved in \eqref{AR1} and \eqref{AR2} and comparing them to \eqref{ToricBlupC}, it turns out that these exact sequences become, in the resolved space, respectively
\be
\begin{tikzcd}[column sep =huge]
\cO(-1) \arrow{r}{\bpp \Sigma_1  \\  - \Sigma_2 \epp} & 
\cO^{\oplus2} \arrow{r}{(\Sigma_1, \Sigma_2)} & \cO(1)\,,
\end{tikzcd}
\ee
and
\be
\begin{tikzcd}[column sep =huge]
\cO(1) \arrow{r}{\bpp \tilde{\Sigma}_1  \\  - \tilde{\Sigma}_2 \epp} & 
\cO^{\oplus2} \arrow{r}{(\tilde{\Sigma}_1, \tilde{\Sigma}_2)} & \cO(-1)\,.
\end{tikzcd}
\ee
Therefore, in the positive phase the upper one remains exact, whereas the lower one fails to be so at the right-most position, giving rise to a sheaf supported on the exceptional $\P^1$. In the negative phase, it is the lower sequence which stays exact, whereas the upper one has a non-trivial cokernel over the flopped $\P^1$. See \cite{Herbst:2008jq} for an account of this transition.

To summarize, using two familiar classes of singularities, we have shown that the sheaves associated to our MF's, whose morphisms we believe describe light membranes hidden in the singularity, are in fact very directly linked to the exceptional $\P^1$'s of the resolved space, which are commonly believed to be wrapped by the heavy membranes at large volume. This provides good evidence for the pertinence of our method. 

Moreover, the sequences \eqref{AR1} and \eqref{AR2} are classified by elements of Ext$^1(M, \tilde M)$, and Ext$^1(\tilde M, M)$, which, by Kn\"orrer's periodicity, are precisely the Ext groups counting open strings between the two D7-branes in the dual IIB picture, as computed in section \ref{4dchirality}. Therefore, this framework relates open strings to vanishing M2-branes quite directly.

\section{Breaking patterns and obstructions to blow-up}\label{BROBsection}

In section \ref{ConifoldSection} we have studied, both in the affine and in the compact case, massless degrees of freedom associated to coherent states of light membranes stuck at a conifold singularity (or a family thereof). In particular, we have seen that, by giving them a vev, we `bind' together two non-Cartier divisors, thus breaking the associated $U(1)$ gauge group. Moreover, the classical geometry remains singular and can no longer be resolved.

In this section we would like to describe the analogous phenomenon for $SU(n)$ singularities. A proper treatment of compact F-theory fibrations with non-abelian gauge groups is beyond the scope of this paper, and will be presented elsewhere. Here, we limit ourselves to the weakly coupled situation, where, ignoring the cubic term in the Weierstrass polynomial, one approximates the elliptic fibration by a multi-Taub-NUT space.  This is already sufficient to discuss in detail how `bound states' of MF's\footnote{Mathematically speaking, we mean extensions between MF's. There are no D-branes in this F/M-theory picture, hence the term `bound state' is only meant as an analogy.} induce the breaking of the original gauge group. In addition, we will propose a criterion to identify the unbroken gauge group, given an MF of the F-theory singular space as input. This criterion is imported from the type IIB context, discussed in \cite{Collinucci:2014qfa}, but it can be equally well applied in F-theory.

In order to study degrees of freedom associated to vanishing M2-branes, we can pass to the \emph{stable category} of MF's of the F-theory internal space, because the ring $R$, which can be thought of as the trivial line bundle, will not contribute to any computations. 
In this category, we can apply Kn\"orrer's periodicity \eqref{upstairs}. By lifting \eqref{suntachyon} we find that the MF of the F-theory internal space corresponding to an \emph{unbroken} $SU(n)$ gauge group has cokernel sheaf given by
\be\label{MFsunstab}
\mathcal{S}_{SU(n)} = \bigoplus_{k=1}^n M_1\quad \Longleftrightarrow\quad (\phi_1,\psi_1)^{\oplus n} \,,
\ee
with 
\be \label{MFstab}
M_i={\rm cokernel} \,(\phi_i\,,\,\psi_i)\,, \quad\quad \phi_i= \bpp u & z^i \\ z^{n-i} & v \epp\,\quad {\rm and} \quad \;\psi_i = \bpp v & -z^i \\ -z^{n-i} & u \epp\,.
\ee
Analogously to the conifold case, turning on vevs for light membranes corresponds to `binding' together the non-Cartier divisors associated to two MF's. In the present context this amounts to switching on non-trivial extensions between the $M_1$'s. A priori, it would seem that such Ext's are empty, since they would be given by vertical maps $\beta$ as follows:
\be
\begin{tikzcd}[row sep=large]
& 0 \rar \dar{\beta} & M_{1}  \\
M_{1} \rar & 0\,.
\end{tikzcd}
\ee
However, in the stable category, one can show that the following exact sequence holds:
\be\label{Mk}
\begin{tikzcd}
0 \arrow{r} & M_{j} \arrow{r} & 
M_{k} \arrow{r} & M_{k-j} \rar & 0\,.
\end{tikzcd}
\ee
for $j, k=1, \ldots, n$.
A special case of this is
\be\label{M1}
\begin{tikzcd}
0 \arrow{r} & M_{k-1} \arrow{r} & 
M_{k} \arrow{r} & M_1 \rar & 0\,.
\end{tikzcd}
\ee
Therefore, $M_1$ can be represented in many ways, thereby giving rise to more morphisms than meet the eye. It turns out that a basis of morphisms that do not factor through other morphisms is given by the following
\be\label{ExtL-L}
\begin{tikzcd}[row sep=large]
& M_k \rar \dar{\beta} & M_{k+1}  \\
M_{k-1} \rar & M_k
\end{tikzcd}
\ee
Since the complexes in \eqref{ExtL-L} are two different representations of $M_1$, this element of ${\rm Ext}^1(M_1, M_1)$ can be associated to the upper triangular MF
\be\label{ExtM1-M1}
\left[\bpp\phi_1 &\beta\\ 0 &\phi_1 \epp\,,\, \bpp\psi_1 &-\alpha\\ 0 &\psi_1 \epp\right]\quad{\rm with}\quad\left\{\begin{array}{l}\phi_1\alpha=\beta\psi_1\,,\\ \psi_1\beta= \alpha\phi_1\,.\end{array}\right.
\ee
The two (non-independent) conditions in \eqref{ExtM1-M1} are just the statement that the pair $(\alpha,\beta)$ is a morphism between the MF's $(\phi_1,\psi_1)$ and $(\psi_1,\phi_1)$, as defined in \eqref{MFmorphisms}. More generally, if elements of Ext$^1$ groups of $p$ consecutive pairs of $M_1$'s are turned on simultaneously, the associated MF will have a $(p+1)\times(p+1)$ Jordan form. These matrices correspond to different breaking patterns of the original gauge group, and the $2n\times2n$ matrix describing the entire system with \emph{broken} $SU(n)$ gauge symmetry will in general be a direct sum of such Jordan blocks. 

Recall from section \ref{ResolutionSection} the definition of the sheaves $L_k$, given by the exact sequences  \eqref{Lk} in the full (not stable) category of MF's.  One can see that if we take elements of the group Ext$^1(L_{k+1},L_k)$, and pass to the stable category \underline{MF}, we get exactly the vertical morphisms in \eqref{ExtL-L}. In section \ref{ResolutionSection} it was also pointed out that the difference of two consecutive $L_k$'s is associated with an exceptional curve $E_k$ of the resolved geometry. Hence, from a group theoretic perspective, this tells us that the fields in the adjoint of $SU(n)$, whose vevs are responsible for the extensions \eqref{ExtM1-M1}, are along the \emph{simple} roots of the original gauge algebra. Moreover, turning them on, besides breaking the gauge group, \emph{obstructs} a certain number of blow-ups, because the fields giving the sizes of some exceptional $\P^1$'s have become massive \cite{Anderson:2013rka}. To identify which blow-ups are still available, it suffices to find all Cartan generators (and linear combinations thereof), which commute with the simple roots turned on. 

It is a well-known fact in representation theory that all gauge inequivalent breaking patterns of this kind (where the rank of the gauge group is lowered) can be distributed in orbits, the so called \emph{nilpotent orbits} \cite{Collingwood}. Such orbits are in one-to-one correspondence with irreducible representations of the Weyl group. Since for $SU(n)$ the Weyl group is the group of permutations of $n$ objects, its nilpotent orbits are in one-to-one correspondence with the integer partitions of $n$. A practical way of classifying them is by using Young tableaux \cite{Collingwood}.\footnote{This technology can also be implemented for $SO(2n)$ groups, if one restricts to a special class of partitions.}
The correspondence between nilpotent orbits and gauge inequivalent extensions of MF's also holds for singularities of the {\bf D} and {\bf E} types, but the details are more involved and will not be presented here.

Since we are working in affine space, we may forget the dependence of all maps on the coordinates which are not involved in the singularity. It is not difficult to see, either by direct computation or by using Kn\"orrer's periodicity, that the conditions in \eqref{ExtM1-M1}, combined with the homotopies of \eqref{ExtL-L}, reduce $\beta$ to
\be
\beta=\lambda\bpp0&1\\ -z^{n-2} &0\epp\,,
\ee
where $\lambda$ is a complex number. Choosing $\lambda=1$, in the same spirit of \eqref{BasisChangeC}, we can make the following left/right independent change of basis
\be
\begin{tikzpicture}[scale=2]
\node (A) at (0,0) {$S^{\oplus 4}$};
\node (B) at (1.6,0)  {$S^{\oplus 4}$};
\node (ar) at (2.9,0) {$\Longrightarrow$};
\node[right = of ar] (C) {$S^{\oplus 4}$};
\node (D) at (5,0) {$S^{\oplus 4}$\,.};
 \draw (A) edge[->] node[above, font=\scriptsize]{$\bpp \phi_1 & \beta \\ 0 & \phi_1 \epp$} (B);
 \draw (C) edge[->] node[above, font=\scriptsize]{$\bpp \phi_0 & 0 \\ 0 & \phi_2 \epp$} (D);
 \draw(A) edge[loop below] node[font=\scriptsize]{$\begin{pmatrix} 1 &0&0& 0\\ 0 & z &0 &1\\ z &0 &-1 &0 \\ 0&1&0&0 \end{pmatrix}$} (A) ;
 \draw(B) edge[loop below] node[font=\scriptsize]{$\begin{pmatrix} 1 &0&0& 0\\ 0 & 0 &0 &1\\ z &0 &-1 &0 \\ 0&1&0&-z \end{pmatrix}$} (B) ;
\end{tikzpicture} 
\ee
After discarding the summand $\phi_0$, which is trivial in the stable category \underline{MF}, we see that, by binding together two copies of $(\phi_1,\psi_1)$, we have generated a copy of $(\phi_2,\psi_2)$, in agreement with \eqref{Mk}. 
Moreover, if we keep binding recursively copies of $M_1$ to our system, we obtain MF's with increasing label. This process ends when we reach $(\phi_n,\psi_n)$, namely when our original MF \eqref{MFsunstab} is reduced to the trivial MF, corresponding to a complete breaking of the original $SU(n)$ gauge symmetry. 

For a general configuration described by a given MF of the ${\bf A}_{\bf n-1}$ singularity, the residual gauge group has the form $S[U(m_1)\cdots U(m_n)]$, where $m_i$ is the multiplicity with which a Jordan block of size $i\times i$ appears in the MF. In the affine case, as we have seen, the number $m_i$ can equivalently be thought of as the multiplicity with which the summand $(\phi_i,\psi_i)$ appears in the MF. However, this does not give us a useful criterion for deciding which gauge group is left unbroken by a general MF of the $SU(n)$ singular F-theory fibration. It is indeed generally not possible to block-diagonalize the given matrix, as we have done in the above toy example in affine space. And in most cases it may also be very hard to isolate Jordan blocks of the type \eqref{ExtM1-M1} or bigger in order to read off their multiplicities.

A possible criterion, which would in principle allow us to handle arbitrary complicated MF's, is based on introducing a `probe' $U(1)$ system and studying the spectrum of charged light membranes transforming in the fundamental of $SU(n)$. The best way to describe it is to go back to type IIB string theory, and describe with a single D7-brane, say on $x=0$, probing a $U(n)$ stack of D7-branes located at $z=0$ by using the tachyon condensation picture. The logic goes as follows: Given a tachyon describing the full system by
\be \label{auxtachyon}
 T = \bpp \phi &0 \\ 0 &x \epp\,,
 \ee
where $\phi$ is has determinant $|\phi| = z^n$. The massless matter spectrum of this system is given by Ext$^1(M,M)$, where $M \equiv$ coker$(T)$ is the cokernel sheaf associated to the tachyon. Focusing on the off-diagonal fluctuations, one finds degrees of freedom with different localization properties. In other words, homotopies restrict them to propagate on loci given by $x=z^i=0$, where the power of $i$ distinguishes the various types of matter. By counting how many fields share the same type of localization, one can deduce the gauge group left unbroken by the $\phi$ component of the tachyon background in \eqref{auxtachyon}. This is because the fundamental representation of the original gauge group breaks up into the sum of fundamentals of the individual factors of the residual group. Therefore the number $m_i$ is the number of fields localized on $x=z^i=0$, and transforming in the fundamental of the factor $U(m_i)$ of the unbroken gauge group.

Now we formulate a criterion for determining the unbroken gauge group directly on the singular F-theory geometry, via analogous reasoning. Given the input MF $(\Phi,\Psi)$ of the polynomial $uv-z^n=0$, we are led to consider the following MF of the auxiliary geometry $uv-xz^n=0$
\be\label{AuxMF_F}
\left[\bpp\Phi' &0 &0 \\ 0 & u &x \\ 0 & z^n & v \epp\,,\, \bpp\Psi' &0 &0 \\ 0 & v &-x \\ 0 & -z^n & u \epp\right]\,.
\ee
The matrices $(\Phi',\Psi')$ are derived from the given ones $(\Phi,\Psi)$ in the following way. One first sets $u=v=0$ in $(\Phi,\Psi)$. According to \eqref{u=v=0}, the MF can be reduced into two direct summands. One now multiplies either one or the other by $x$ and restores the variables $u$ and $v$ by applying \eqref{upstairs}. By computing Ext$^1(\hat{M},\hat{M})$, where $\hat{M}$ is the cokernel sheaf associated to the MF \eqref{AuxMF_F}, one can study the localization properties of the off-diagonal fluctuations, which, as we have learned, are associated to light membranes. The numbers $m_i$ are finally obtained by counting how many degrees of freedom are localized on the loci $u=v=x=z^i=0$, for $i=1,\ldots,n$.

It is possible to formulate a criterion which does not rely on having a probe $U(1)$ system. This is particularly useful when dealing with compact F-theory models, where Kn\"orrer's periodicity is generally not available, and one does not want to modify or spoil the original singularity in order to have an additional $U(1)$. Given the input MF $(\Phi,\Psi)$ of the F-theory geometry, one studies the localization properties of the \emph{adjoint} massless spectrum by computing Ext$^1({\rm coker}(\Phi,\Psi),{\rm coker}(\Phi,\Psi))$. At this point, one can reason in terms of the multiplicities of degrees of freedom with equal localization properties, as we have done before, in order to deduce the unbroken gauge group.

\section{Discussion}\label{ConclSection}

In this paper we have initiated a program which aims to address certain fundamental questions on the physics of singular F-theory compactifications without the need for removing the singularity. For this purpose, we have exploited the powerful tool of matrix factorizations in a variety of familiar situations,
focusing, in particular, on two basic questions:
\begin{itemize}
\item Compute the charged spectrum, counting chiral and anti-chiral fields separately;
\item Explore backgrounds of the theory that are inaccessible by any means that rely on removing the singularity.
\end{itemize}
We have always made our computations directly on the F-theory space, and used the corresponding weakly coupled type IIB picture only for the purpose of comparing results. The success of this comparison as well as the deep connections of our method with the geometry of the resolutions give us strong confidence that the technique developed here could be equally well employed in situations which lack a well-behaved weakly coupled description.

This new framework opens up new avenues of investigation, raising several interesting questions:
\begin{itemize}

\item In this paper, we have focused on gauge groups that can be obtained perturbatively. In that case, given an unbroken gauge group, there is a canonical choice among all possible MF's of the CY fourfold. However, for general gauge groups this correspondence needs to be clarified. 

\item A related problem is that of finding the appropriate globally defined MF's for ADE singularities on elliptic fibrations.

\item We worked in the holomorphic gauge. D-term constraints need to be incorporated in this picture in order to fully understand the effective theory of an F-theory background.
\item To our knowledge, the only other approach to dealing with F-theory on singular spaces is the proposal of using \emph{limiting mixed Hodge structures} of \cite{Anderson:2013rka}. It would be interesting to explore possible connections between that and our approach.

\item The set of matrix factorizations of a singularity naturally comes equipped with a quiver, the \emph{Auslander-Reiten quiver}, discussed in \ref{ResolutionSection}. One reasonable expectation is that this quiver is implicitly describing some gauge theory, presumably of a probe of the F/M-theory singularity. It would be interesting to figure this out.

\end{itemize}

\section*{Acknowledgements}
We are grateful for the interesting and inspiring discussions we had with Lara Anderson, Riccardo Argurio, Ilka Brunner, Cyril Closset, Pierre Corvilain, Eran Palti, Marco Fazzi, Hirotaka Hayashi, Johnathan Heckman, Michael Kay, Wolfgang Lerche, Ruben Minasian, Dave Morrison.
A. C. would like to give special thanks to Andreas Braun and Roberto Valandro for countless discussions during which some precursors of the ideas presented here were born.
 
 A. C. is a Research Associate of the Fonds de la Recherche Scientifique F.N.R.S. (Belgium). The work of A. C. is partially supported by IISN - Belgium (convention 4.4514.08).
The work of R. S. was supported by the ERC Starting Independent Researcher Grant 259133-ObservableString.

R. S. would like to thank the Universit\'e Libre de Bruxelles for hospitality during part of this project. The authors are grateful to the Mainz Institute for Theoretical Physics (MITP) for its hospitality and its partial support during the completion of this work. A. C. and R. S. gratefully acknowledge support from the Simons Center for Geometry and Physics, Stony Brook University at which some of the research for this paper was performed.

\providecommand{\href}[2]{#2}\begingroup\raggedright\endgroup


\begin{thebibliography}{10}

\bibitem{Vafa:1996xn}
C.~Vafa, ``{Evidence for F theory},''
  \href{http://dx.doi.org/10.1016/0550-3213(96)00172-1}{{\em Nucl.Phys.} {\bf
  B469} (1996)  403--418},
\href{http://arxiv.org/abs/hep-th/9602022}{{\tt arXiv:hep-th/9602022
  [hep-th]}}.

\bibitem{Witten:1995ex}
E.~Witten, ``{String theory dynamics in various dimensions},''
  \href{http://dx.doi.org/10.1016/0550-3213(95)00158-O}{{\em Nucl. Phys.} {\bf
  B443} (1995)  85--126},
\href{http://arxiv.org/abs/hep-th/9503124}{{\tt arXiv:hep-th/9503124
  [hep-th]}}.

\bibitem{Bershadsky:1995sp}
M.~Bershadsky, C.~Vafa, and V.~Sadov, ``{D strings on D manifolds},''
  \href{http://dx.doi.org/10.1016/0550-3213(96)00024-7}{{\em Nucl. Phys.} {\bf
  B463} (1996)  398--414},
\href{http://arxiv.org/abs/hep-th/9510225}{{\tt arXiv:hep-th/9510225
  [hep-th]}}.

\bibitem{Bershadsky:1996nh}
M.~Bershadsky, K.~A. Intriligator, S.~Kachru, D.~R. Morrison, V.~Sadov, {\em et
  al.}, ``{Geometric singularities and enhanced gauge symmetries},''
  \href{http://dx.doi.org/10.1016/S0550-3213(96)90131-5}{{\em Nucl.Phys.} {\bf
  B481} (1996)  215--252},
\href{http://arxiv.org/abs/hep-th/9605200}{{\tt arXiv:hep-th/9605200
  [hep-th]}}.

\bibitem{Strominger:1995cz}
A.~Strominger, ``{Massless black holes and conifolds in string theory},''
  \href{http://dx.doi.org/10.1016/0550-3213(95)00287-3}{{\em Nucl.Phys.} {\bf
  B451} (1995)  96--108},
\href{http://arxiv.org/abs/hep-th/9504090}{{\tt arXiv:hep-th/9504090
  [hep-th]}}.

\bibitem{Bies:2014sra}
M.~Bies, C.~Mayrhofer, C.~Pehle, and T.~Weigand, ``{Chow groups, Deligne
  cohomology and massless matter in F-theory},''
\href{http://arxiv.org/abs/1402.5144}{{\tt arXiv:1402.5144 [hep-th]}}.

\bibitem{Grassi:2013kha}
A.~Grassi, J.~Halverson, and J.~L. Shaneson, ``{Matter From Geometry Without
  Resolution},'' \href{http://dx.doi.org/10.1007/JHEP10(2013)205}{{\em JHEP}
  {\bf 1310} (2013)  205},
\href{http://arxiv.org/abs/1306.1832}{{\tt arXiv:1306.1832 [hep-th]}}.

\bibitem{Grassi:2014sda}
A.~Grassi, J.~Halverson, and J.~L. Shaneson, ``{Non-Abelian Gauge Symmetry and
  the Higgs Mechanism in F-theory},''
\href{http://arxiv.org/abs/1402.5962}{{\tt arXiv:1402.5962 [hep-th]}}.

\bibitem{Cecotti:2010bp}
S.~Cecotti, C.~Cordova, J.~J. Heckman, and C.~Vafa, ``{T-Branes and
  Monodromy},'' \href{http://dx.doi.org/10.1007/JHEP07(2011)030}{{\em JHEP}
  {\bf 1107} (2011)  030},
\href{http://arxiv.org/abs/1010.5780}{{\tt arXiv:1010.5780 [hep-th]}}.

\bibitem{Donagi:2011jy}
R.~Donagi and M.~Wijnholt, ``{Gluing Branes, I},''
  \href{http://dx.doi.org/10.1007/JHEP05(2013)068}{{\em JHEP} {\bf 1305} (2013)
   068},
\href{http://arxiv.org/abs/1104.2610}{{\tt arXiv:1104.2610 [hep-th]}}.

\bibitem{Anderson:2013rka}
L.~B. Anderson, J.~J. Heckman, and S.~Katz, ``{T-Branes and Geometry},''
  \href{http://dx.doi.org/10.1007/JHEP05(2014)080}{{\em JHEP} {\bf 1405} (2014)
   080},
\href{http://arxiv.org/abs/1310.1931}{{\tt arXiv:1310.1931 [hep-th]}}.

\bibitem{Eisenbud:1980}
D.~Eisenbud, ``{Homological algebra on a complete intersection, with an
  application to group representations},'' {\em Trans. Amer. Math. Soc.} {\bf
  260} (1980)  35--64.

\bibitem{Douglas:1996sw}
M.~R. Douglas and G.~W. Moore, ``{D-branes, quivers, and ALE instantons},''
\href{http://arxiv.org/abs/hep-th/9603167}{{\tt arXiv:hep-th/9603167
  [hep-th]}}.

\bibitem{Berenstein:2001jr}
D.~Berenstein and R.~G. Leigh, ``{Resolution of stringy singularities by
  noncommutative algebras},''
  \href{http://dx.doi.org/10.1088/1126-6708/2001/06/030}{{\em JHEP} {\bf 0106}
  (2001)  030},
\href{http://arxiv.org/abs/hep-th/0105229}{{\tt arXiv:hep-th/0105229
  [hep-th]}}.

\bibitem{Bergh:aa}
M.~V. den Bergh, ``Non-commutative crepant resolutions,''
  \href{http://arxiv.org/abs/math/0211064}{{\tt math/0211064}}.
  \url{http://arxiv.org/abs/math/0211064}.

\bibitem{Aspinwall:2010mw}
P.~S. Aspinwall and D.~R. Morrison, ``{Quivers from Matrix Factorizations},''
  \href{http://dx.doi.org/10.1007/s00220-012-1520-1}{{\em Commun.Math.Phys.}
  {\bf 313} (2012)  607--633},
\href{http://arxiv.org/abs/1005.1042}{{\tt arXiv:1005.1042 [hep-th]}}.

\bibitem{Esole:2011sm}
M.~Esole and S.-T. Yau, ``{Small resolutions of SU(5)-models in F-theory},''
  \href{http://dx.doi.org/10.4310/ATMP.2013.v17.n6.a1}{{\em
  Adv.Theor.Math.Phys.} {\bf 17} (2013)  1195--1253},
\href{http://arxiv.org/abs/1107.0733}{{\tt arXiv:1107.0733 [hep-th]}}.

\bibitem{Diaconescu:1998ua}
D.-E. Diaconescu and S.~Gukov, ``{Three-dimensional N=2 gauge theories and
  degenerations of Calabi-Yau four folds},''
  \href{http://dx.doi.org/10.1016/S0550-3213(98)00597-5}{{\em Nucl.Phys.} {\bf
  B535} (1998)  171--196},
\href{http://arxiv.org/abs/hep-th/9804059}{{\tt arXiv:hep-th/9804059
  [hep-th]}}.

\bibitem{Grimm:2011fx}
T.~W. Grimm and H.~Hayashi, ``{F-theory fluxes, Chirality and Chern-Simons
  theories},'' \href{http://dx.doi.org/10.1007/JHEP03(2012)027}{{\em JHEP} {\bf
  1203} (2012)  027},
\href{http://arxiv.org/abs/1111.1232}{{\tt arXiv:1111.1232 [hep-th]}}.

\bibitem{Cvetic:2012xn}
M.~Cvetic, T.~W. Grimm, and D.~Klevers, ``{Anomaly Cancellation And Abelian
  Gauge Symmetries In F-theory},''
  \href{http://dx.doi.org/10.1007/JHEP02(2013)101}{{\em JHEP} {\bf 1302} (2013)
   101},
\href{http://arxiv.org/abs/1210.6034}{{\tt arXiv:1210.6034 [hep-th]}}.

\bibitem{Hayashi:2013lra}
H.~Hayashi, C.~Lawrie, and S.~Schafer-Nameki, ``{Phases, Flops and F-theory:
  SU(5) Gauge Theories},''
  \href{http://dx.doi.org/10.1007/JHEP10(2013)046}{{\em JHEP} {\bf 1310} (2013)
   046},
\href{http://arxiv.org/abs/1304.1678}{{\tt arXiv:1304.1678 [hep-th]}}.

\bibitem{Hayashi:2014kca}
H.~Hayashi, C.~Lawrie, D.~R. Morrison, and S.~Schafer-Nameki, ``{Box Graphs and
  Singular Fibers},'' \href{http://dx.doi.org/10.1007/JHEP05(2014)048}{{\em
  JHEP} {\bf 05} (2014)  048},
\href{http://arxiv.org/abs/1402.2653}{{\tt arXiv:1402.2653 [hep-th]}}.

\bibitem{Kapustin:2002bi}
A.~Kapustin and Y.~Li, ``{D branes in Landau-Ginzburg models and algebraic
  geometry},'' \href{http://dx.doi.org/10.1088/1126-6708/2003/12/005}{{\em
  JHEP} {\bf 0312} (2003)  005},
\href{http://arxiv.org/abs/hep-th/0210296}{{\tt arXiv:hep-th/0210296
  [hep-th]}}.

\bibitem{Hori:2004zd}
K.~Hori and J.~Walcher, ``{D-branes from matrix factorizations},''
  \href{http://dx.doi.org/10.1016/j.crhy.2004.09.016}{{\em Comptes Rendus
  Physique} {\bf 5} (2004)  1061--1070},
\href{http://arxiv.org/abs/hep-th/0409204}{{\tt arXiv:hep-th/0409204
  [hep-th]}}.

\bibitem{Brunner:2003dc}
I.~Brunner, M.~Herbst, W.~Lerche, and B.~Scheuner, ``{Landau-Ginzburg
  realization of open string TFT},''
  \href{http://dx.doi.org/10.1088/1126-6708/2006/11/043}{{\em JHEP} {\bf 0611}
  (2006)  043},
\href{http://arxiv.org/abs/hep-th/0305133}{{\tt arXiv:hep-th/0305133
  [hep-th]}}.

\bibitem{Jockers:2007ng}
H.~Jockers and W.~Lerche, ``{Matrix Factorizations, D-Branes and their
  Deformations},''
  \href{http://dx.doi.org/10.1016/j.nuclphysbps.2007.06.012}{{\em
  Nucl.Phys.Proc.Suppl.} {\bf 171} (2007)  196--214},
\href{http://arxiv.org/abs/0708.0157}{{\tt arXiv:0708.0157 [hep-th]}}.

\bibitem{Braun:2011zm}
A.~P. Braun, A.~Collinucci, and R.~Valandro, ``{G-flux in F-theory and
  algebraic cycles},''
  \href{http://dx.doi.org/10.1016/j.nuclphysb.2011.10.034}{{\em Nucl.Phys.}
  {\bf B856} (2012)  129--179},
\href{http://arxiv.org/abs/1107.5337}{{\tt arXiv:1107.5337 [hep-th]}}.

\bibitem{Grimm:2010ez}
T.~W. Grimm and T.~Weigand, ``{On Abelian Gauge Symmetries and Proton Decay in
  Global F-theory GUTs},''
  \href{http://dx.doi.org/10.1103/PhysRevD.82.086009}{{\em Phys.Rev.} {\bf D82}
  (2010)  086009},
\href{http://arxiv.org/abs/1006.0226}{{\tt arXiv:1006.0226 [hep-th]}}.

\bibitem{Yoshino}
Y.~Yoshino, {\em Maximal Cohen-Macaulay Modules Over Cohen-Macaulay Rings}.
\newblock Cambridge University Press, 1990.

\bibitem{MCMbook}
R.~W. Graham~Leuschke, {\em Cohen-Macaulay representations}, vol.~181 of {\em
  Mathematical Surveys and Monographs}.
\newblock AMS, 2012.

\bibitem{Herbst:2008jq}
M.~Herbst, K.~Hori, and D.~Page, ``{Phases Of N=2 Theories In 1+1 Dimensions
  With Boundary},''
\href{http://arxiv.org/abs/0803.2045}{{\tt arXiv:0803.2045 [hep-th]}}.

\bibitem{Collinucci:2014qfa}
A.~Collinucci and R.~Savelli, ``{T-branes as branes within branes},''
\href{http://arxiv.org/abs/1410.4178}{{\tt arXiv:1410.4178 [hep-th]}}.

\bibitem{Sen:1998sm}
A.~Sen, ``{Tachyon condensation on the brane anti-brane system},''
  \href{http://dx.doi.org/10.1088/1126-6708/1998/08/012}{{\em JHEP} {\bf 9808}
  (1998)  012},
\href{http://arxiv.org/abs/hep-th/9805170}{{\tt arXiv:hep-th/9805170
  [hep-th]}}.

\bibitem{Knorrer}
H.~Kn\"orrer, ``Cohen-macaulay modules on hypersurface singularities i,'' {\em
  Invent. Math.} {\bf 88} (1987)  153--164.

\bibitem{Clingher:2012rg}
A.~Clingher, R.~Donagi, and M.~Wijnholt, ``{The Sen Limit},''
\href{http://arxiv.org/abs/1212.4505}{{\tt arXiv:1212.4505 [hep-th]}}.

\bibitem{Donagi:2012ts}
R.~Donagi, S.~Katz, and M.~Wijnholt, ``{Weak Coupling, Degeneration and Log
  Calabi-Yau Spaces},''
\href{http://arxiv.org/abs/1212.0553}{{\tt arXiv:1212.0553 [hep-th]}}.

\bibitem{Braun:2014nva}
A.~P. Braun, A.~Collinucci, and R.~Valandro, ``{The fate of U(1)'s at strong
  coupling in F-theory},''
  \href{http://dx.doi.org/10.1007/JHEP07(2014)028}{{\em JHEP} {\bf 1407} (2014)
   028},
\href{http://arxiv.org/abs/1402.4054}{{\tt arXiv:1402.4054 [hep-th]}}.

\bibitem{Katz:2002gh}
S.~H. Katz and E.~Sharpe, ``{D-branes, open string vertex operators, and Ext
  groups},'' {\em Adv.Theor.Math.Phys.} {\bf 6} (2003)  979--1030,
\href{http://arxiv.org/abs/hep-th/0208104}{{\tt arXiv:hep-th/0208104
  [hep-th]}}.

\bibitem{Freed:1999vc}
D.~S. Freed and E.~Witten, ``{Anomalies in string theory with D-branes},'' {\em
  Asian J.Math} {\bf 3} (1999)  819,
\href{http://arxiv.org/abs/hep-th/9907189}{{\tt arXiv:hep-th/9907189
  [hep-th]}}.

\bibitem{Sen:1996vd}
A.~Sen, ``{F theory and orientifolds},''
  \href{http://dx.doi.org/10.1016/0550-3213(96)00347-1}{{\em Nucl.Phys.} {\bf
  B475} (1996)  562--578},
\href{http://arxiv.org/abs/hep-th/9605150}{{\tt arXiv:hep-th/9605150
  [hep-th]}}.

\bibitem{Braun:2014pva}
A.~P. Braun, A.~Collinucci, and R.~Valandro, ``{Hypercharge flux in F-theory
  and the stable Sen limit},''
  \href{http://dx.doi.org/10.1007/JHEP07(2014)121}{{\em JHEP} {\bf 1407} (2014)
   121},
\href{http://arxiv.org/abs/1402.4096}{{\tt arXiv:1402.4096 [hep-th]}}.

\bibitem{Collinucci:2008pf}
A.~Collinucci, F.~Denef, and M.~Esole, ``{D-brane Deconstructions in IIB
  Orientifolds},'' \href{http://dx.doi.org/10.1088/1126-6708/2009/02/005}{{\em
  JHEP} {\bf 0902} (2009)  005},
\href{http://arxiv.org/abs/0805.1573}{{\tt arXiv:0805.1573 [hep-th]}}.

\bibitem{Witten:2009xu}
E.~Witten, ``{Branes, Instantons, And Taub-NUT Spaces},''
  \href{http://dx.doi.org/10.1088/1126-6708/2009/06/067}{{\em JHEP} {\bf 0906}
  (2009)  067},
\href{http://arxiv.org/abs/0902.0948}{{\tt arXiv:0902.0948 [hep-th]}}.

\bibitem{Wemyss:2012ee}
M.~Wemyss, ``{Lectures on Noncommutative Resolutions},''
\href{http://arxiv.org/abs/1210.2564}{{\tt arXiv:1210.2564 [math.RT]}}.

\bibitem{Collingwood}
D.~H. Collingwood and W.~M. McGovern, {\em Nilpotent orbits in semisimple Lie
  algebras}.
\newblock Van Nostrand, 1993.

\end{thebibliography}
\end{document}